\documentclass[10pt,journal,compsoc]{IEEEtran}

\usepackage[utf8]{inputenc}
\usepackage{enumitem}
\usepackage{amsmath}
\usepackage[normalem]{ulem}
\usepackage{verbatim}
\usepackage{algorithm, algorithmicx} 
\usepackage{algcompatible}
\usepackage{wrapfig}

\usepackage{url}
\usepackage{soul}
\usepackage{amsmath, amssymb, amsfonts}
\usepackage{textcomp}
\usepackage{siunitx}
\usepackage{xcolor,colortbl}
\usepackage{subfig}
\usepackage{tcolorbox}
\usepackage{float}
\usepackage{graphicx}  
\usepackage{xspace}

\newcommand{\cm}[1]{{\bf\color{red}[#1]}}
\newcommand{\cb}[1]{{\bf\color{black}#1}}
\newcommand{\mypara}[1]{\smallskip\noindent{\bf {#1}}~}

\newcommand{\sfrtxt}[1]{\textcolor{black}{#1}}

\newcommand{\system}{\texttt{DynoLoc}\xspace}

\newenvironment{packeditemize}{
	\begin{list}{$\bullet$}{
			\setlength{\itemsep}{1.5pt}
			\setlength{\labelwidth}{8pt}
			\setlength{\leftmargin}{10pt}
			\setlength{\labelsep}{3pt}
			\setlength{\listparindent}{\parindent}
			\setlength{\parsep}{1.5pt}
			\setlength{\parskip}{1.5pt}
			\setlength{\topsep}{1.5pt}}}{\end{list}
			}

\makeatletter
\def\endthebibliography{%
	\def\@noitemerr{\@latex@warning{Empty `thebibliography' environment}}%
	\endlist
}
\makeatother

\begin{document}
\title{{\system}: Infrastructure-free RF Tracking in Dynamic Indoor Environments\vspace{-0ex}}


%
\author{
	\IEEEauthorblockN{Md. Shaifur Rahman\IEEEauthorrefmark{1},  Ayon Chakraborty\IEEEauthorrefmark{2}, Karthikeyan Sunderasan\IEEEauthorrefmark{3}, Sampath Rangarajan\IEEEauthorrefmark{4}\\}\vspace{10pt}
	
	\footnotesize{The work was done when all the authors were employees of NEC Laboratories America and is protected by the published patent applications: US20210306977A1~\cite{pat1} and US20210185491A1~\cite{pat2}. The current affiliations of the authors are mentioned in the footnote below.}

}
\IEEEtitleabstractindextext{%
	\begin{abstract}

Promising solutions exist today that can accurately track mobile entities indoor using visual inertial odometry in {\em favorable visual conditions}, or by leveraging fine-grained ranging (RF, ultrasonic, IR, etc.) to {\em reference anchors}.
However,  they are unable to directly cater to ``{\em dynamic}'' indoor environments (e.g. first responder scenarios, multi-player AR/VR gaming in everyday spaces, etc.) that are devoid of such favorable conditions. 
Indeed, we show that the need for ``infrastructure-free'', and robustness to ``node mobility'' and ``visual conditions'' in such environments, motivates a robust RF-based approach along with the need to address a novel and challenging 
variant of its  
infrastructure-free (i.e. peer-to-peer) localization problem that is {\em latency-bounded} -- accurate tracking of mobile entities  imposes a latency budget that not only affects the solution computation but also the collection of peer-to-peer ranges themselves.


In this work, we present the design and deployment of \system that addresses this latency-bounded infrastructure-free RF localization problem. To this end, \system unravels the fundamental tradeoff between latency and localization accuracy and incorporates design elements that judiciously leverage the available ranging resources to adaptively estimate the {\em joint} topology of nodes, coupled with robust algorithm that maximizes the localization accuracy even in the face of practical environmental artifacts (wireless connectivity and multipath, node mobility, etc.). This allows \system to track (every second) a network of few tens of mobile entities even at speeds of 1-2 m/s with median accuracies under 1-2\,m (compared to 5m+ with baselines), without infrastructure support. We demonstrate \system's potential in a real-world firefighters' drill, as well as two other use cases of (i) multi-player AR/VR  gaming, and (ii) active shooter tracking by first responders.   
	\end{abstract}

}

\maketitle

\footnote{	\IEEEauthorrefmark{1}Stony Brook University
	\IEEEauthorrefmark{2}IIT Madras
	\IEEEauthorrefmark{3}Georgia Tech
	\IEEEauthorrefmark{4}Peraton Labs
}
\section{Introduction} 
%

\begin{figure}[!htb]
	\centering
		\includegraphics[width=0.8\linewidth]{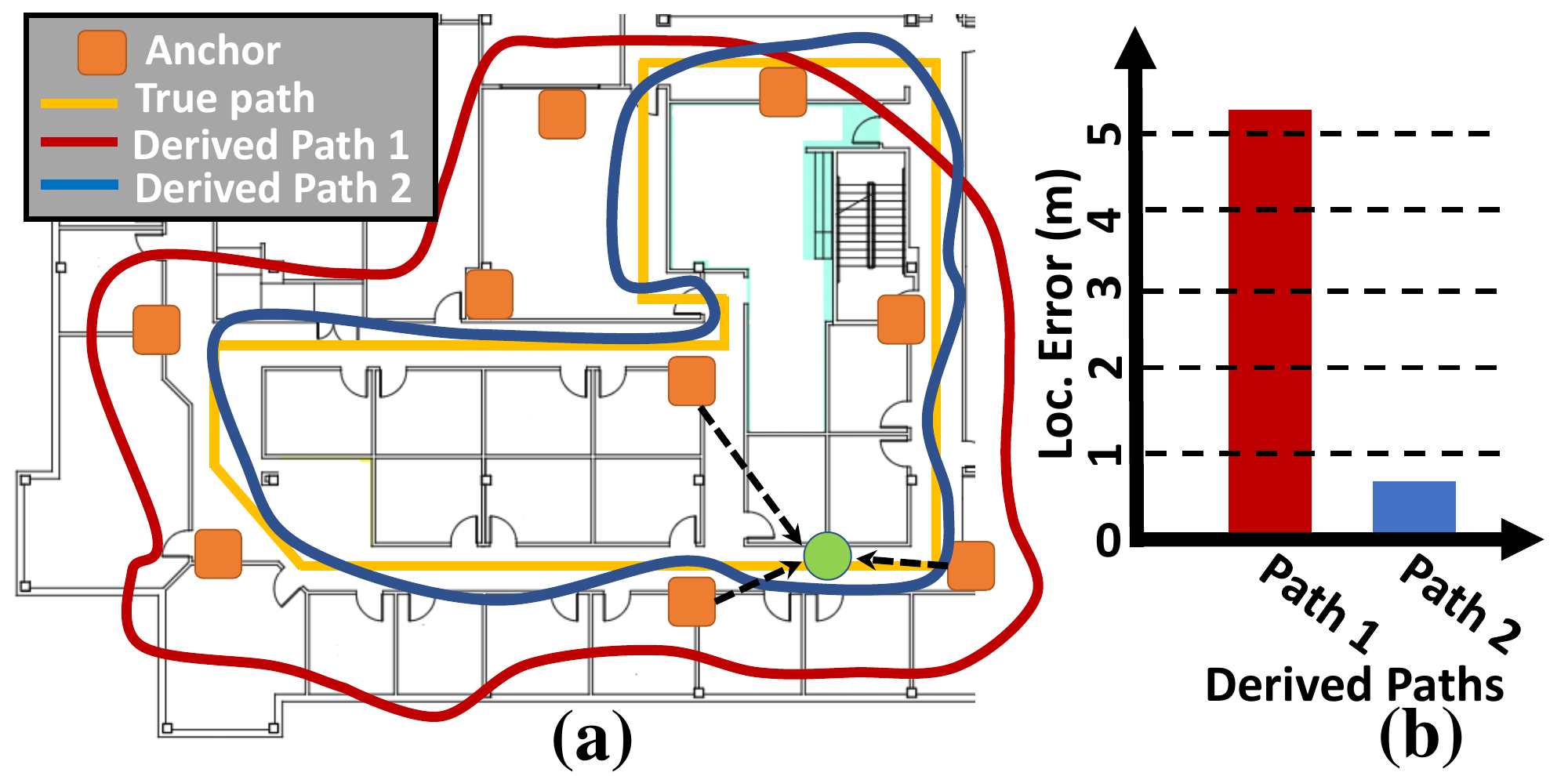}
	\caption{
	 \sfrtxt{(a) Two paths on the floormap dotted by fixed anchors- the mobile node on i) path 1 ranges with all anchors ii) on path 2, ranges with nearest/LOS 3 anchors (b) Avg. localization errors for the 2 derived paths}
	}
	\label{fig:latency-bound-op}
\end{figure}

\begin{figure}[!htb]
	\centering
		\includegraphics[width=0.98\linewidth]{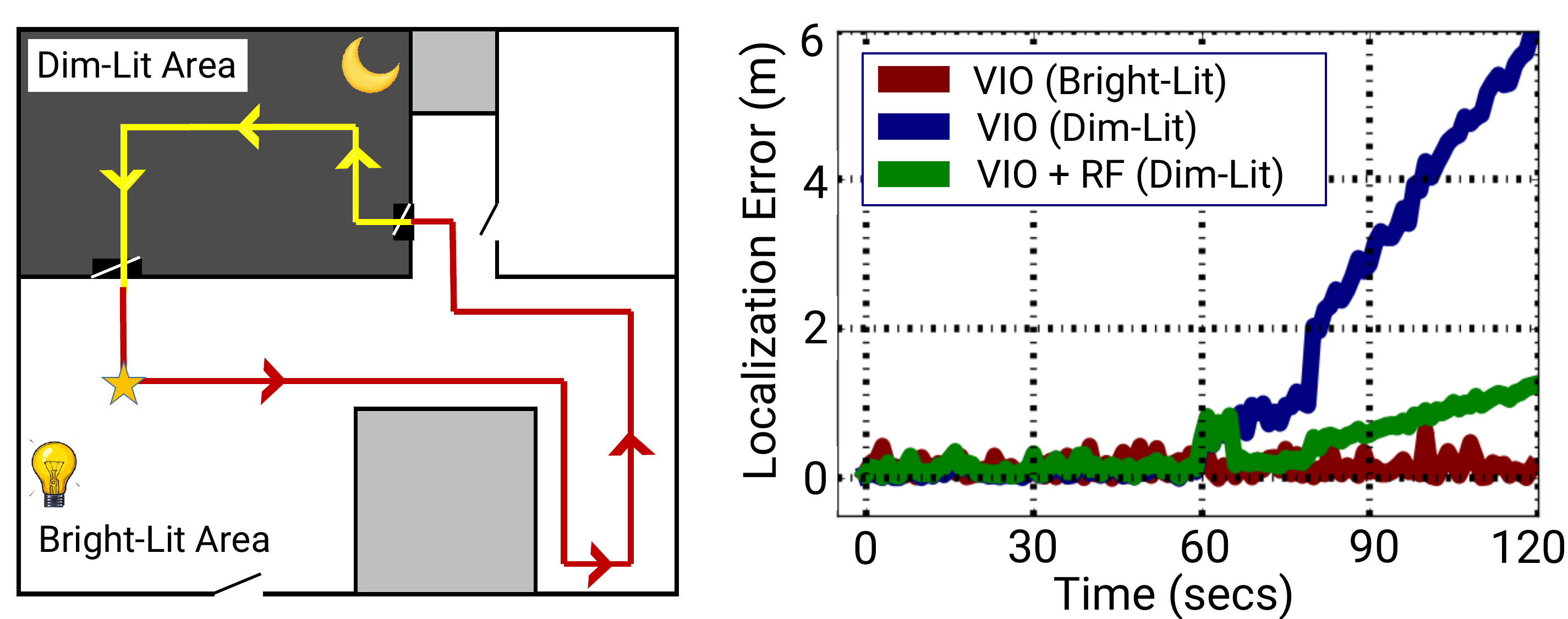}
	\caption{
	 A simple experiment to demonstrate persistent localization errors in a VIO based system (Intel RealSense~\cite{intel-realsense}). VIO exhibits high tracking accuracy under good lighting conditions, but deteriorates greatly when traversing a dim-lit region, from which it is unable to recover. Fusing RF-based (UWB anchors) localization with VIO can curtail such accumulation of errors.
	}
	\label{fig:vio-indoor-exp}
\end{figure}

{\bf Dynamic indoor environments.}
Several promising solutions exist for indoor localization today that leverage various modalities (RF~\cite{spotfi-15, chronos-16, md-track-19, array-track-13, spotfi-15}, ultrasonic~\cite{ultrasound-loc-survey}, optical (IR) tracking~\cite{htc-vive-benchmark}, etc.) and multiple dimensions (antennas, channels, access points, etc.) to provide fine-grained (sub-meter, decimeter-level) localization.
A natural  question that arises is {\em do we really need another  indoor localization solution?}
The answer indeed depends on the capabilities of the environment, where the solution is deployed. 
The bulk of today's localization solutions 
rely on the deployment of static anchor nodes (in known locations) that provide distance estimation to a target client (a.k.a. ranging), which are then aggregated to deliver its location. 
\sfrtxt{However, ranging is a latency-bound operation and localization without taking into consideration the cost of operation, node mobility, link quality (LOS/NLOS) can cause poor accuracy in localization even with fixed anchors, see path-1 vs. path-2 in Fig. \ref{fig:latency-bound-op} which are derived by without and with consideration of the above factors for a mobile node.
}
On the other hand, infra-free solutions based on inertial sensors are prone to accumulating errors of tens of meters over time~\cite{bo2013smartloc,zhang2012inertial}; Even the high-accuracy (cm-level) solutions for AR/VR that are based on Visual Inertial Odometry (VIO)~\cite{vio-dev-survey} (fuse cameras and IMUs), suffer appreciably (as shown in Fig.~\ref{fig:vio-indoor-exp}) when the environment is not well-lit/textured, has motion blur and/or dynamic entities in screen, etc.  
Unfortunately, these existing anchor-based and infra-free solutions do not lend themselves to ``dynamic" indoor environments that are inherently {\em characterized by the lack of reference anchors, unfavorable visual conditions, and mobile entities}. 


\begin{figure*}
	\centering
	\includegraphics[width = 1\linewidth]{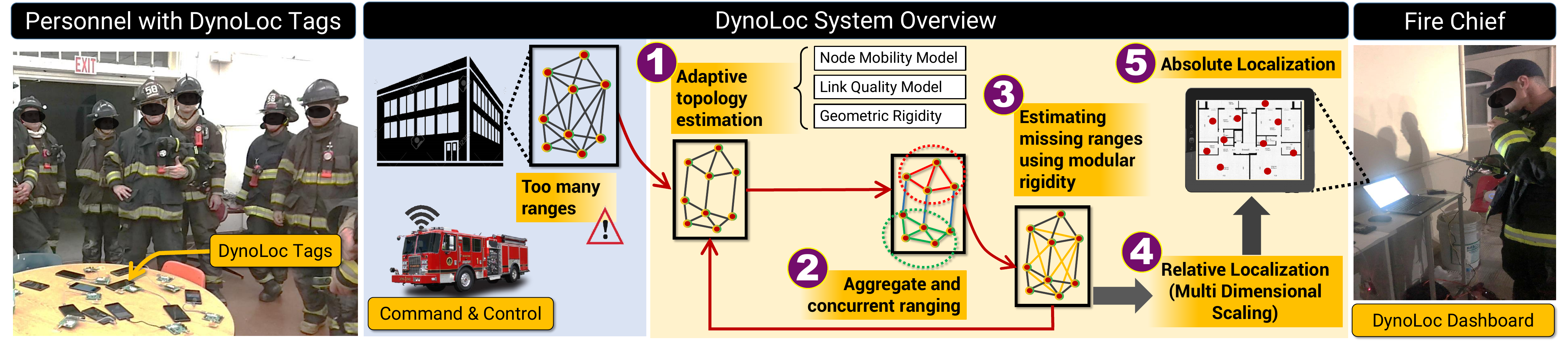}
	\caption{\system was deployed and tested in a real firefighters' drill attended by 10$+$ firefighters in a 2-storey building (100\,m$\times$50\,m). The basic building blocks of \system's algorithm are shown as a schematic that estimates absolute locations of the firefighters. The fire chief at the {\em command and control} station gets a real-time visual feedback through our dashboard application.}
	\vspace{-10 pt}
	\label{fig:dynoloc-overview}
	\vspace{-5 pt}
\end{figure*} 

\noindent {\bf RF-based localization for dynamic environments:}
Localization and tracking in such dynamic environments, especially with mobile clients, are not only central to all first responder scenarios, but also enable new capabilities in emerging consumer applications like mixed-reality (MR) gaming,   where 
multiple players engaged in an AR/VR game can be freely tracked in real-time across large, everyday (unmapped) indoor spaces that span multiple rooms in less-than-favorable visual conditions. An RF-based localization solution that can deliver high accuracies (sub-m if not cm) without relying on anchors, can fill this critical need as a stand-alone solution for first responder applications, and a complementary solution to VIO (for alleviating its errors) for MR gaming applications, owing to its robustness in unfavorable visual conditions.  
%

\noindent{\bf Gap between ranging and localization.}
 Obtaining accurate ranges (i.e. distances) between the clients and anchors in infrastructure solutions, automatically leads to accurate  localization of clients (also referred to as nodes) through multilateration. However, in the absence of such anchor infrastructure, nodes are capable of ranging only with respect to each other. There exists a large technical gap in going from such relative ranging to localization in  dynamic environments, where existing multilateration approaches cannot be leveraged.  

\noindent{\bf Challenges in addressing the gap.}
Localizing nodes in an absolute frame of reference may be challenging without one or more reference nodes. 
However, existing works in the sensor literature
have shown that if one can estimate the relative geometry of nodes (called {\em relative localization},  e.g. Fig.~\ref{fig:rigid}(a))  using their pair-wise measured ranges; then, additional information (such as IMU data, floor plan, etc.) can be used to potentially rotate, translate or flip this relative geometry to obtain the {\em absolute localization} of the nodes.
The resulting efforts that focused on relative localization~
albeit amenable to theoretical analysis, however, do not account for a critical dimension needed for practical deployments, namely {\em node mobility}. Incorporating the latter however, significantly changes the nature of the problem, requiring one to solve the {\em latency-bounded} version of the infrastructure-free RF localization problem, which has not been addressed before. Indeed, infrastructure-free localization solutions today are unable to track even a network of around 10 mobile nodes (at just 1 m/s speed) with an accuracy of under 6m (as shown later in Fig.~\ref{fig:base-var-percent-mobile}). This can be attributed to the following key challenges:

%
\noindent{\em (i) Latency vs. accuracy tradeoff:} Accurate location of the nodes needs to be tracked at least every second (i.e. refresh rate of 1 Hz) 
for a node mobility of 1-1.5 m/s. The corresponding latency constraint restricts the number of  node-pairs that can be ranged (before computing a localization solution), thereby 
lowering the accuracy of localization significantly by several folds (Fig. \ref{fig:var-refresh-rate}). Further, the quality of links (edges) and hence the ranges measured, are in turn impacted by the  geometry of the induced topology, node mobility as well as the multi-path wireless channel, and have a large impact on the accuracy as well (Figs.~\ref{fig:var-noisy-link},~\ref{fig:var-mobile-node}).

\noindent{\em (ii) Overhead of range measurements:}
Ranging between node pairs is typically accomplished through sequential packet exchanges and time-of-flight estimation techniques
, thereby incurring a large latency and hence reduced ability to track a large network of mobile nodes (Fig.~\ref{fig:var-node-count}) . 

\noindent{\em (iii) Partial information degrades accuracy:}
Existing solutions for relative localization (e.g. techniques using Euclidean distance matrices, EDM
) work well when network topologies are a complete graph and all range estimates are available and accurate. However, in the absence of such features in practical deployments, the accuracy can suffer appreciably (Figs. \ref{fig:var-topologies},~\ref{fig:var-configurations}).


\noindent{\bf $\blacksquare$ \system (Dynamic Indoor Localization) design.} Towards addressing these challenges, we present \system -- a system for latency-bounded infrastructure-free localization that can be readily deployed in  dynamic indoor environments. While \system's framework is agnostic to the underlying wireless technology (e.g. WiFi, UWB, mmWave) used for ranging, it currently employs UWB, given the latter's ability to offer good ranging resolution (tens of cm) at reasonable indoor penetration (70-90m LOS, 30-50m NLOS).
\system equips each of the nodes that need to be tracked with a tag that encompasses a UWB radio (for ranging), WiFi radio (for control/orchestration), and IMU. While UWBs are the primary source of active ranging, IMUs are used in a limited scope (heading and mobility indication) only to resolve  ambiguities in localization. \system's design involves three key components:

\noindent\underline{\em Topology estimation for ranging:}
\system intelligently uses its available ranging resources on {\em critical} links that will contribute the most to topology's  localization accuracy.
The critical nature of a link varies spatio-temporally and is determined by \system by fusing three dimensions of information, namely (a) mobility of nodes in the link (that affect the staleness of its range measurement),  (b) certainty of range estimates being LOS vs. NLOS (inferred from channel impulse response measurements), and (c) link's contribution to the topology's geometry in creating a robust and maximally rigid (where relative location of nodes are fixed in the topology) sub-graph that in turn leads to increased localization accuracy. 

\noindent\underline{\em Aggregated and concurrent ranging:} 
\system redesigns the traditional pair-wise and sequential ranging protocol for reduced measurement latency. It aggregates (and amortizes), the process of ranging  (and associated overhead) for a node with all its neighbors into a single compacted process, while links that are spatially separated, can enable such ranging concurrently.  

\noindent\underline{\em Robust relative localization:}
Instead of applying EDM-completion techniques on the entire topology that is incomplete, \system leverages the graph rigidity construct of {\em k-core} sub-graphs to identify maximal rigid sub-graphs of the topology, and applies EDM on these separately and combines them to provide a robust, accurate solution. 
Given the relative localization of nodes in the rigid sub-graphs, \system devises additional mechanisms to localize the remaining nodes in the topology by leveraging  geometric constraints driven by range, physical connectivity as well as heading data from IMUs. 
 
Finally, with little additional meta information, contributed by IMU heading data or floor plans, \system efficiently transforms the relative localization solution into absolute coordinate system without affecting the solution's refresh rate.

\noindent{\bf $\blacksquare$ \system's potential.} We have built and deployed \system in real-world dynamic environments, including in a live firefighters' drill (Fig.~\ref{fig:dynoloc-overview}), where its accuracy and value in saving lives was well-appreciated (Sec.~\ref{subsec:in-the-wild}). Designed with mobility (and hence latency) in mind, 
\system delivers superior performance in infra-free tracking across multiple dimensions of node density, mobility, application refresh rate, etc. In particular,  evaluation in two real-world use-cases, reveal that (i) {\em first-responder scenarios:} \system is able to track a network of 12 (20) responders, operating with speeds upto 1 m/s  (2 m/s) with a median localization error of under  1 m (2 m), while delivering a refresh rate of 1 Hz; existing solutions suffer in accuracy (6m+ error) even for a node mobility of 1 m/s; (ii) {\em AR gaming:} \system tracks translational motion of users accurately across a free-flowing indoor space of 20m$\times{}$20m  with latencies of under 64 msec to enable a highly-responsive, dynamic 3-player AR game 
even in sub-optimal lighting conditions - a scenario that is challenging for VIO, especially in the multi-player context. 
In future, we aim to fuse VIO with \system to deliver on VIO's performance even in realistic everyday scenarios.
A short video demonstrating \system's capabilities is available at the following link:\\\cb{\texttt{http://sites.google.com/view/dynoloc}} 

Our contributions in this work are two fold.
\begin{packeditemize}
\item{} We introduce and address the problem of {\em latency-bounded, infrastructure-free localization} that is critical for tracking in several dynamic indoor applications.  

\item{} We build and demonstrate the viability of such a system called \system{}, in two real world use-cases targeting first responders (a real firefighters' drill and active shooter tracking) and multi-player AR/VR gaming. 
\end{packeditemize}
%
\section{Motivation \& Challenges}
%
\label{sec:background}
We consider the problem of tracking mobile entities (nodes) in {\em dynamic} indoor environments, namely those that are {\em un-calibrated, lack externally deployed localization infrastructure, and characterized by node mobility}. 
This features a pressing need in first responder situations (as evident from NIST programs~\cite{nist-survey}), with the potential to enable anchor-less user tracking in multi-player AR/VR gaming applications in the future. 
  
%
\begin{figure}[!thb]
	\centering
	\includegraphics[width = 0.99\linewidth ]{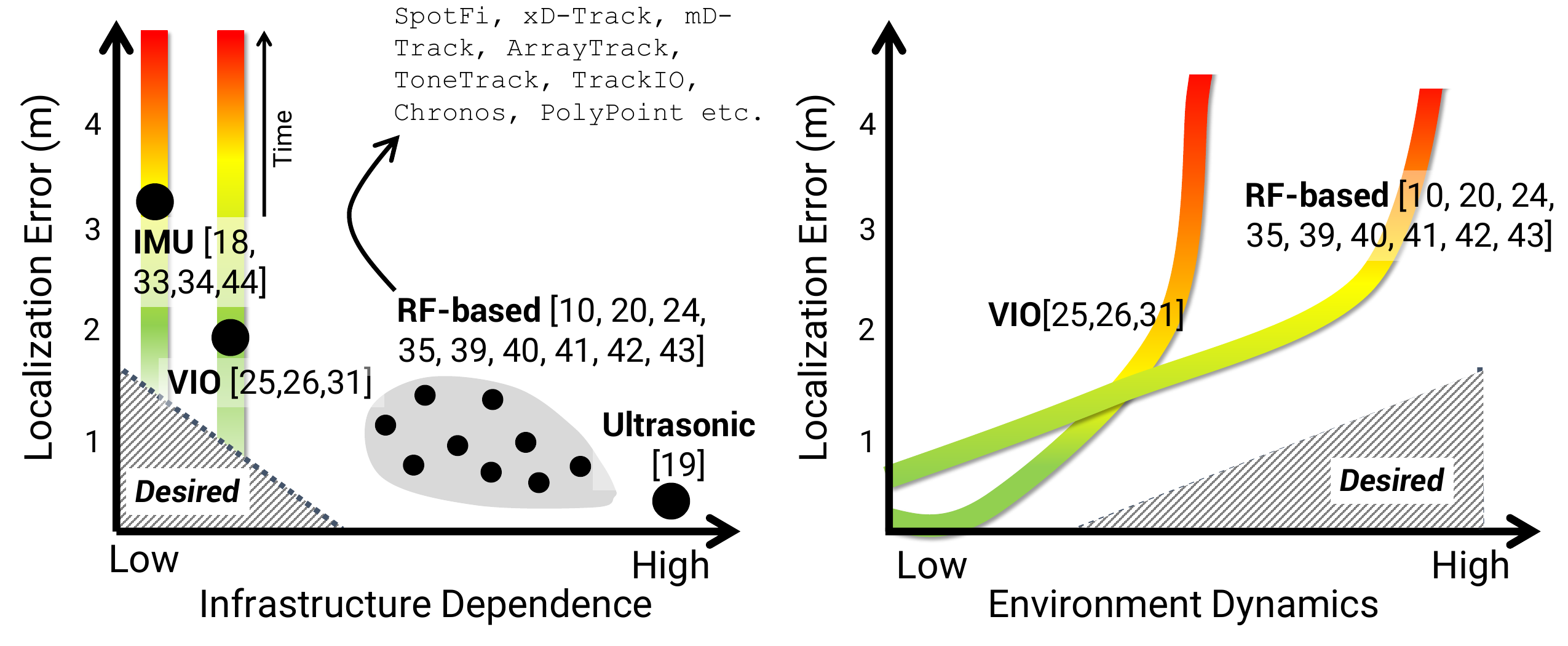}
	\caption{Localization accuracy is shown as a function of dependence on deployed infrastructure ({\em left}) and dynamics of the environment ({\em right})}
	\label{fig:lit-trade-off}
\end{figure}
\subsection{Background on Related Works}
The rich literature in the area of active (locating and identifying) indoor localization
can be broadly categorized as (i) Anchor-based, and (ii) Infrastructure-free approaches, as shown in Fig.~\ref{fig:lit-trade-off}.

\mypara{Anchor-based:}
These approaches often surpass their infra-free counterparts in accuracy 
 at the expense of a-priori deployed infrastructure for localization -- a tradeoff captured in  Fig.~\ref{fig:lit-trade-off}.
%
Here, beacons are deployed at known locations and serve as reference points or anchors. A node estimates its distances (also called ranges) from 
three such anchors, which are then combined with the anchors' locations to estimate its own location by a technique typically known as multilateration. Given a technology to perform accurate ranging, localization can be seen as a {\em trivial} extension. Hence, most of the prior works in this space has focused on the accurate estimation of such ranges, 
particularly using WiFi access points as beacons, while some have also leveraged ultrasonic beacons~\cite{ultrasound-loc-survey}. The WiFi-based works  leverage signal information across multiple dimensions -- frequency (~\cite{tonetrack-15}, ~\cite{chronos-16}), antenna arrays (~\cite{spotfi-15}, ~\cite{monoloco-18}), or both (~\cite{md-track-19}), to improve accuracy in the face of limited WiFi bandwidth and multipath. Some of them~\cite{locap} adopt a finger-printing approach (using RSSI, CSI, etc.) to calibrate the environment a-priori that is later used for for real-time location inference.  Optical tracking systems (e.g. HTC Vive~\cite{htc-vive-benchmark}) that are popular in the AR/VR industry, employ multiple IR beacons (LEDs/ cameras) to provide mm-level tracking accuracy, but are restricted to line-of-sight and expensive to deploy. 
%

The fundamental dependence on pre-deployed anchors (mostly static, but sometimes mobile -- e.g. outdoor drones~\cite{dhekne2019trackio}), prevents such approaches from catering to our target environment.  
%
%

\mypara{Infrastructure-free:}
Works in this category are more amenable to our target environment, but exhibit a different tradeoff between accuracy and robustness, as captured in Fig.~\ref{fig:lit-trade-off}. 
Inertial sensor-based solutions~\cite{bo2013smartloc, zhang2012inertial, romit-imu-16, romit-imu-18} are inherently local to a node (no ranging needed), and hence popular.
However, with only dead-reckoning of nodes, errors accumulate significantly over time~\cite{intertial-loc-survey}, especially in case of pedestrian mobility.
State-of-the-art AR/VR solutions (e.g. ARCore~\cite{arcore-cap}) leverage visual inertial odometry (VIO) that combines both cameras and IMUs to provide accurate cm-level tracking~\cite{vio-dev-survey} in favorable conditions. However, they often require anchors \cite{rowe-ipin-19} and their performance suffers significantly in poorly illuminated and/or poorly textured environments~\cite{opto-survey}, in presence of motion-blur \cite{opto-blur} and/or multiple moving objects in the video-frames \cite{opto-obj}, as shown in Fig.~\ref{fig:vio-indoor-exp}.  Such practical conditions result in various errors relating to drift, loop-closure, scale ambiguity etc. (for SLAM-based approaches) \cite{vslam-ov}, and errors related to projection, parameterization etc. (for optical-flow based approaches) \cite{opt-flow-ov}.
\\
\sfrtxt{
	\cite{lifi-14} uses CSI of WiFi to determine LOS. \cite{p2ploc-18} uses P2P UWB ranging for localization of mobile node considering dilution of precision. \cite{calib-free-ipsn-2017} uses Gaussian Mixture and MDS to solve indoor localization using UWB ranges.
}
\\
Hence, 
dynamic environments, particularly those in first responder scenarios can significantly benefit from an alternate RF modality that can deliver good accuracies (sub-1-2m), and which is robust to the lacking of such favorable conditions. 
In 
AR/VR gaming applications, such a modality can be complementary in helping to eliminate the accumulating errors faced by VIO, with periodic absolute location fixes.
\subsection{Role of RF in Infra-free Localization:}
The recent popularity of ultra wide-band (UWB~\cite{decawave}) technology, and its ability to span a wide 500-1000 MHz bandwidth with superior multipath suppression (owing to its impulse transmissions), has made it a popular candidate for sub-m localization~\cite{warehouseWSN,fernandez2007application,gowda2017bringing, uwb_pos}, albeit with the help of infrastructure anchors. 
Existing works in this space are largely concerned with scalable ranging (SurePoint~\cite{kempke2016surepoint}, SnapLoc~\cite{grobetawindhager2019snaploc}) and tracking of individual mobile nodes (e.g., indoor drone, PolyPoint~\cite{kempke2015polypoint}). 
However,  in the absence of reference anchors, UWB's two-way-ranging (TWR~\cite{sahinoglu2006ranging}) mechanism can enable the nodes to only range with each other. Hence, localization in  dynamic environments presents a different 
 challenge, which, beyond the estimation of accurate ranges, needs to translate the ranges to an accurate localization solution.
Indeed, the key focus of infra-free RF localization comes down to bridging this gap between estimated ranges and node localization that arises in the absence of anchors.        

\mypara{Primer on Relative Localization:}
In contrast to anchor-based approaches, where nodes are {\em absolutely} localized  in the coordinate space defined by the anchors, localization in infra-free set-up is a two-step process. First, the nodes are {\em relatively} localized among themselves, following which  some meta information (e.g. orientation of the nodes, or floorplans) is leveraged to transform such relative localization to the absolute coordinate space. Relative localization refers to the geometry or a topology among the nodes, where the pairwise distance between nodes as well as their relative orientation are preserved. 
In determining such a relative localization, the construct of a {\em rigid body} comes handy. 

\begin{figure}[!thb]
	\centering
	\includegraphics[width = 0.7\linewidth ]{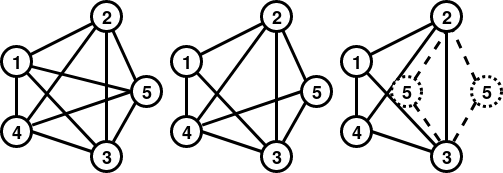}
	\caption{(a) Clique (rigid) (b) non-clique (rigid) (c) 2 choices for node 5 (not rigid) }
	\label{fig:rigid}
\end{figure}

Let $G=(V, E, d)$ be a weighted graph, where $d$ is the set of range measurements (weights) for the |E| edges defined on the |V| nodes.
A \textit{realization} is a function $x:V \rightarrow{} \mathbb{R}^2$ that maps the set of vertices $V$ to the 2D Euclidean space such that each range value is preserved i.e $\forall (u,v) \in E, \; ||x(u) - x(v)|| = d(u, v)$ where $||.||$ is the Euclidean norm. The graph $G=(V, E, d)$ is ``rigid" if there is only one realization, discounting any translation, rotation and flip. Thus, a rigid topology gives us a unique relative localization solution. As shown in the Figure \ref{fig:rigid}, the complete graph or clique in (a) is rigid because given the 10 ranges between all  pairs of nodes, this graph is the only realization, although it can be rotated, translated and/or flipped. Similarly, given the 9 ranges, the graph in (b) is rigid. However, (c) is not rigid, since given the 8 ranges, there are two choices to fix node 5 relative to the edge (2,3) resulting in two potential realizations.

\begin{figure}[!htb]
	\centering
	\subfloat[]
	{
		\includegraphics[width=0.4\linewidth]{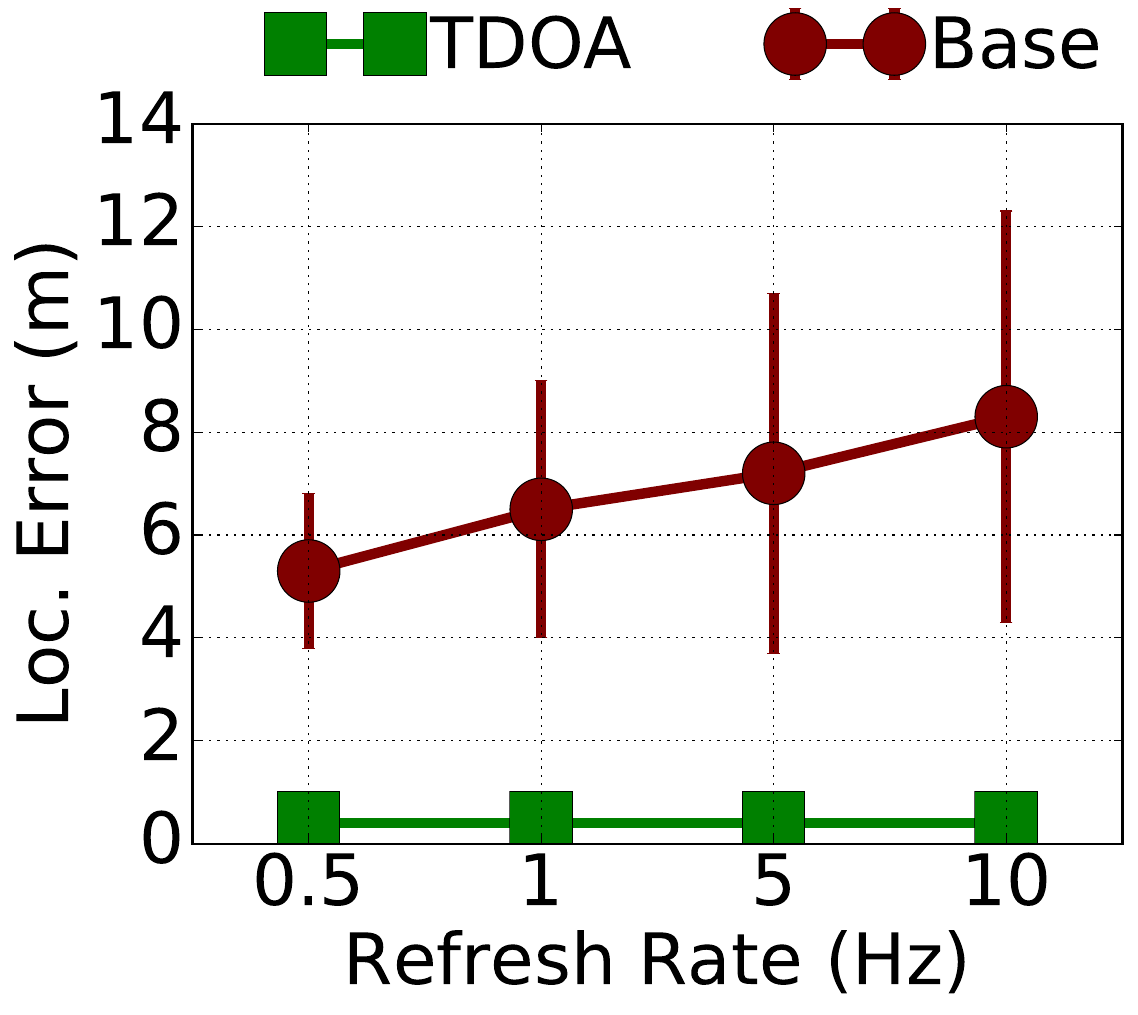}
		\label{fig:var-refresh-rate}
	}
	\subfloat[]
	{
		\includegraphics[width=0.4\linewidth]{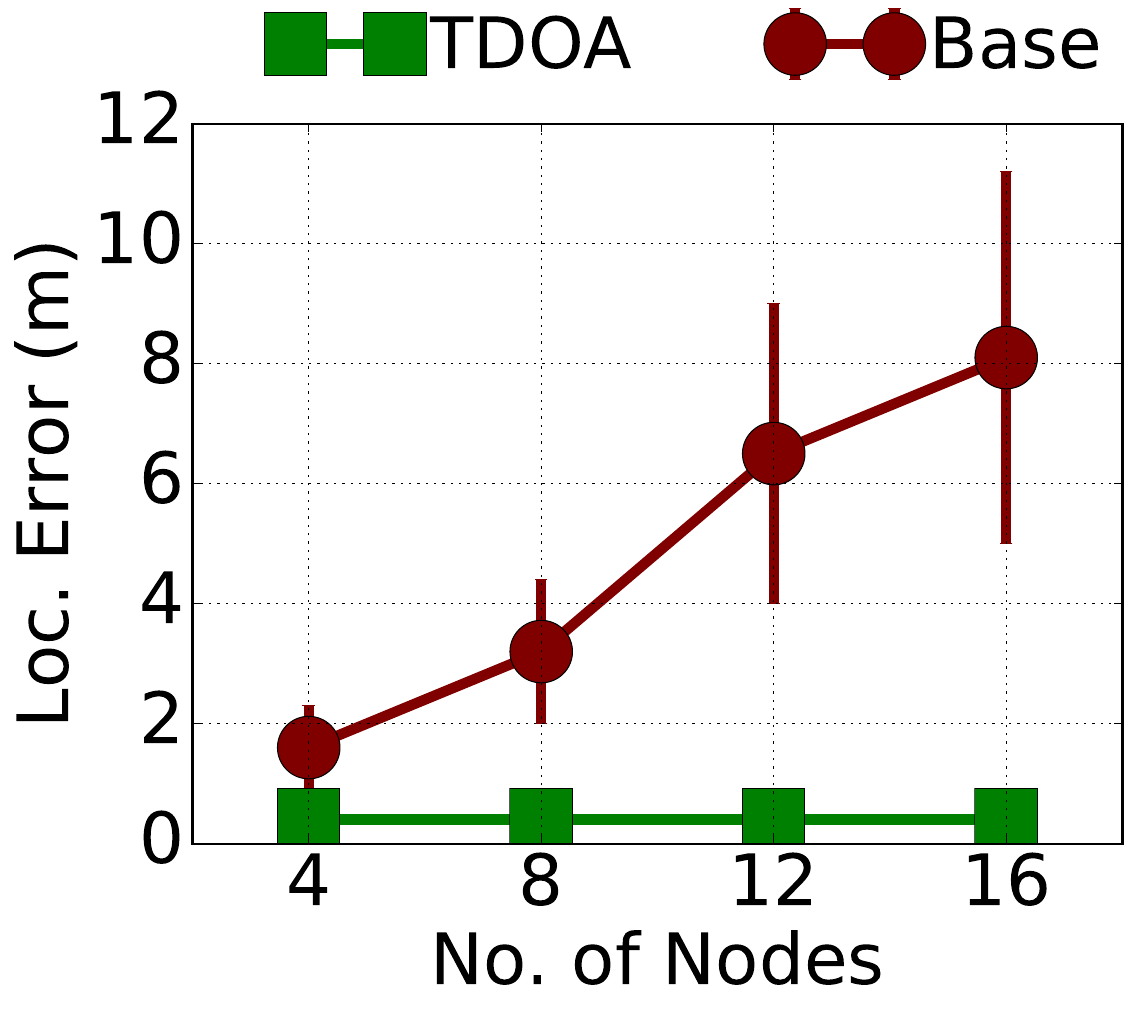}
		\label{fig:var-node-count}
	}
	\vspace{-10 pt}
	\caption{Localization error for 
		(a) various refresh rates 
		(b) various no. of nodes 
	}
	\vspace{-7 pt}
\end{figure}
\begin{figure}[!htb]
	\centering
	\subfloat[]
	{
		\includegraphics[width=0.4\linewidth]{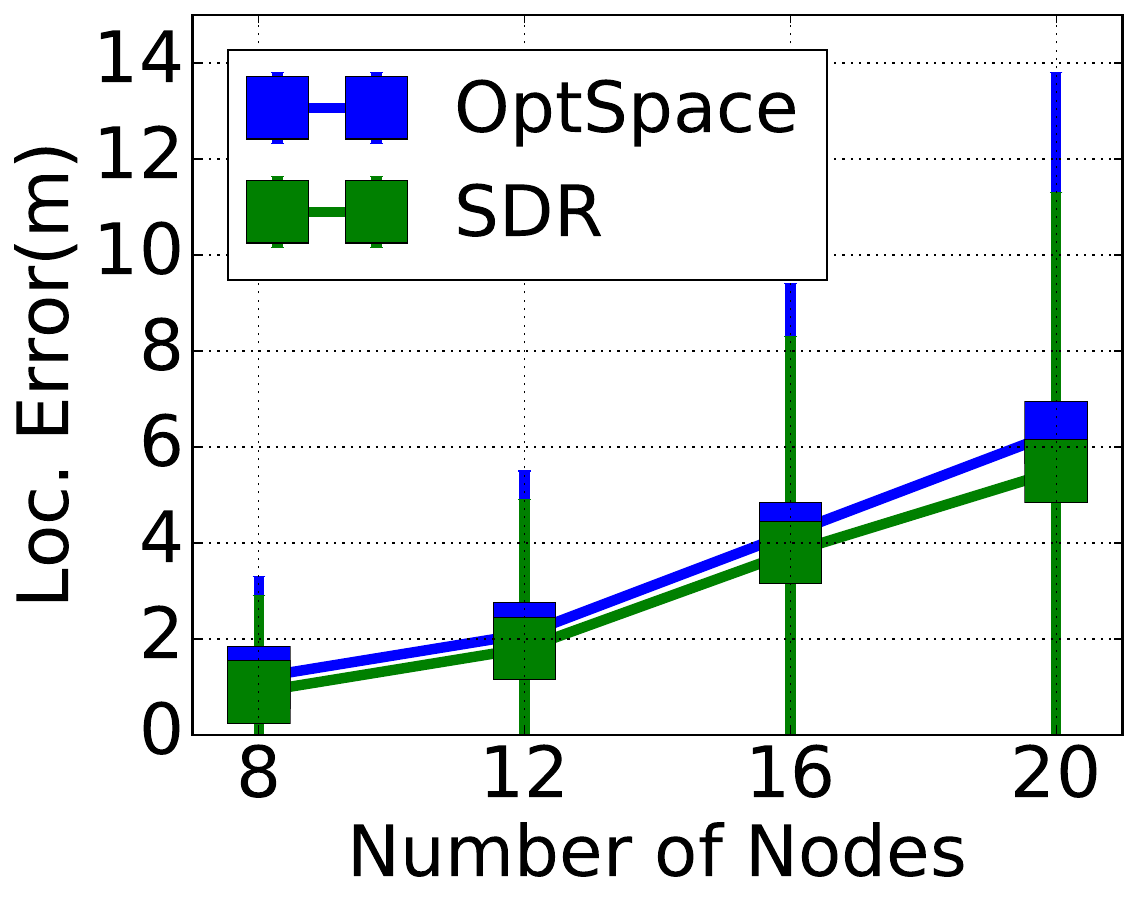}
		\label{fig:var-topologies}
	}
	\subfloat[]
	{
		\includegraphics[width=0.4\linewidth]{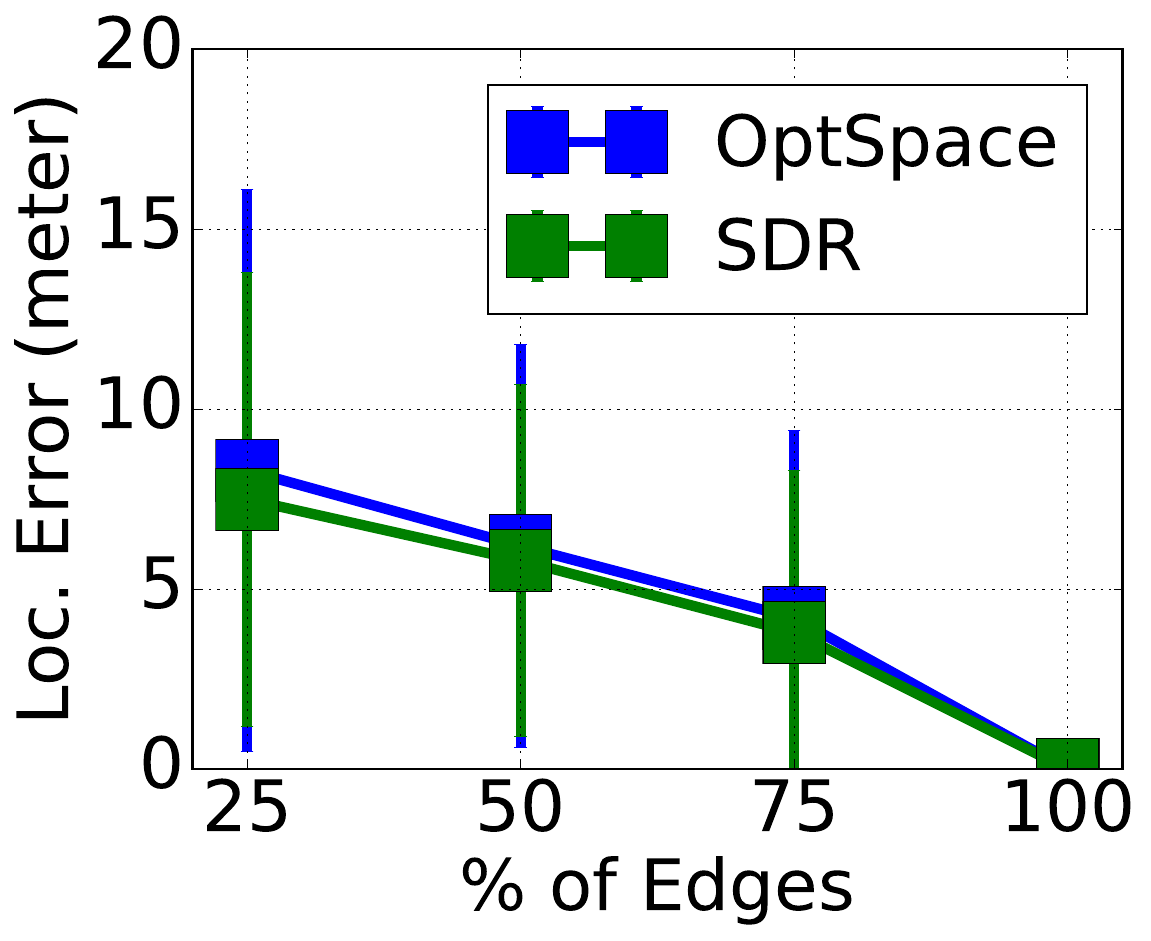}
		\label{fig:var-configurations}
	}
	\vspace{-10 pt}
	\caption{Localization error after using various low-rank matrix completion techniques for(a) various topologies and (b) various set of edges from same topology 
	}
	\vspace{-13 pt}
\end{figure}
\begin{figure}[!htb]
	\centering
	\subfloat[]
	{
		\includegraphics[width=0.4\linewidth]{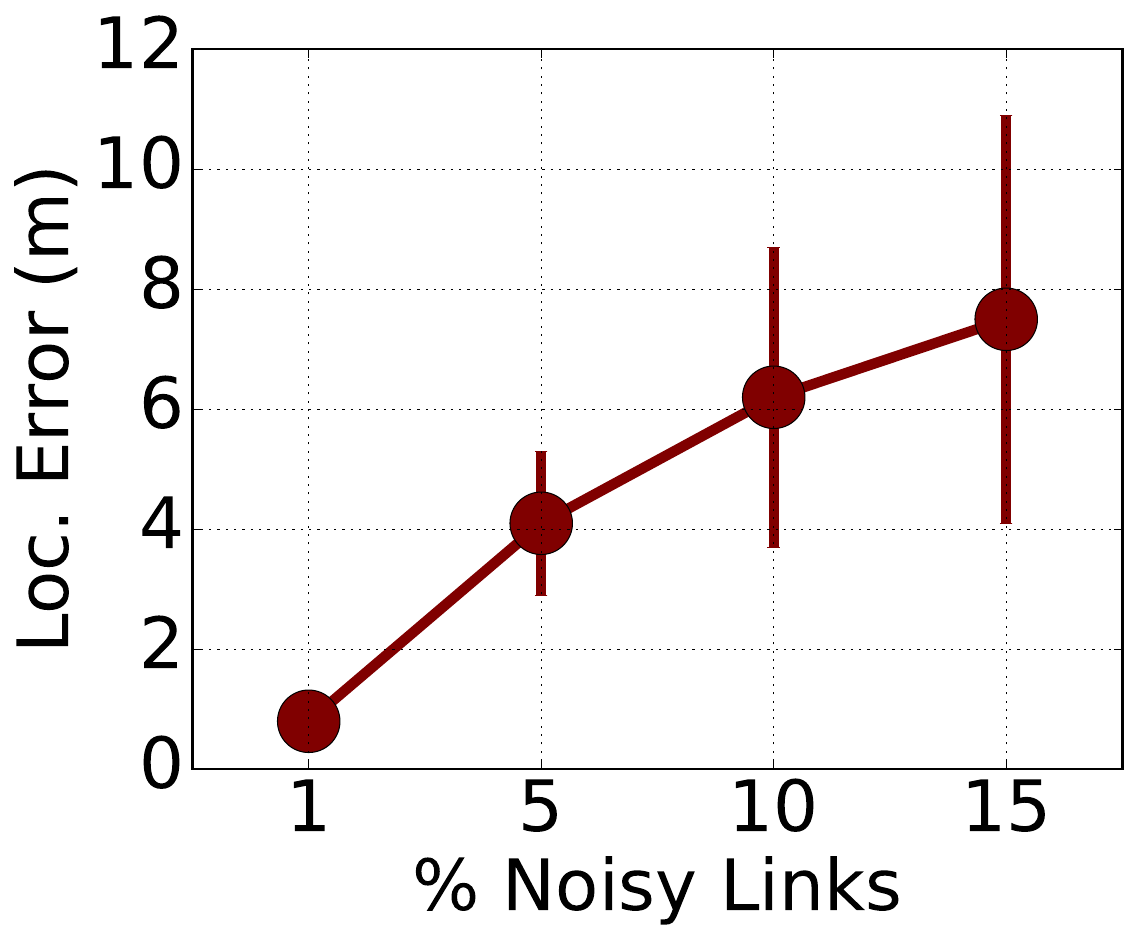}
		\label{fig:var-noisy-link}
	}
	\subfloat[]
	{
		\includegraphics[width=0.4\linewidth]{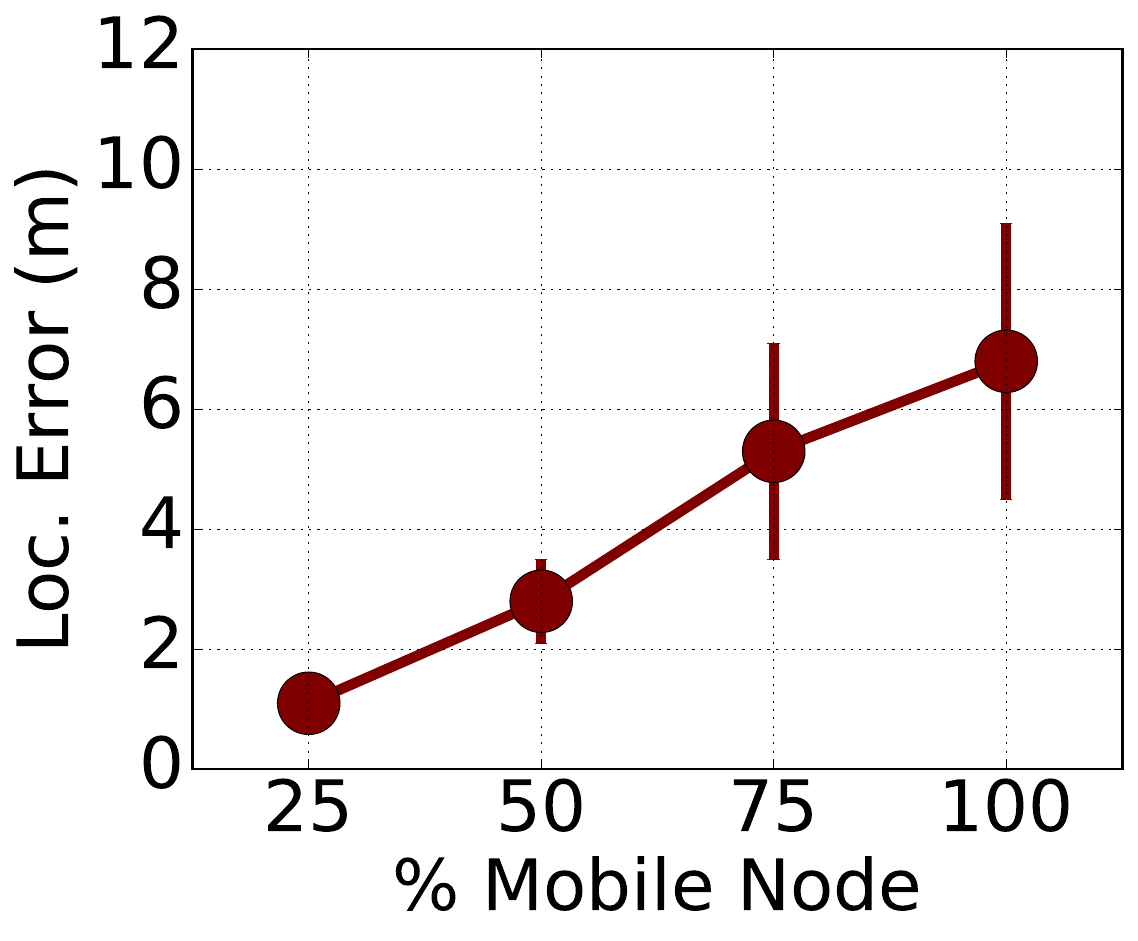}
		\label{fig:var-mobile-node}
	}
	\caption{Localization error for
		(a) various levels of link noises
		(b) various fractions of mobile nodes}
\end{figure}

If {\em all} possible ranges between nodes are available, computing the relative localization is straight-forward in a {\em static} environment. The edge weights of the graph are maintained in the form of an adjacency matrix, \texttt{EDM} (a.k.a. Euclidean Distance Matrix), where each entry represents a measured range between two nodes. Then, an approach called Multidimensional Scaling (MDS) is applied on the \texttt{EDM} matrix, whereby an EVD (Eigen Value Decomposition) results in an embedding of the nodes (i..e. relative localization) in a 2D Cartesian space. 
While such a framework of relative localization is appropriate for our dynamic environments, 
the theoretical approaches in this space are built on several assumptions\footnote{e.g. static nodes, accurate range estimates of all node pairs at no cost, etc.} that do not hold in practice.
\subsection{Challenges in a Dynamic Environment}
\label{subsec:challenges}
To accurately localize/track mobile nodes, the localization solution needs to be computed and refreshed at a granularity finer than node mobility. For instance, a 1 Hz refresh rate is appropriate to track nodes with speeds of 1-1.5 m/s, targeting a 1-2 m error. 
However,  the refresh rate automatically enforces a latency-bound (cost) for the whole process of relative localization, which involves both the range estimation/collection as well as the solution computation. This results in a {\em latency-bounded} version of the infra-free localization problem (referred to as LB-IFL) that has not been addressed before.    
To understand the impact of such a latency cost on existing approaches in practice, we conduct an experimental study (details of testbed described in Section 4) comparing a genie/anchor-aided localization solution (with all possible range estimates available instantaneously, TDOA) with the one described above, i.e., MDS applied on EDM constructed with a random set of edges (called Base), whose ranges are measured within the latency budget offered by the refresh rate.   


\mypara{Latency vs. accuracy:}
Every ranging operation takes a finite amount of time to complete. For instance, with a popular UWB hardware~\cite{decawave}, it takes approximately 40\,ms to complete a single range estimate, i.e. a two-way-ranging (TWR) operation. This inherently limits the number of ranges that can be estimated/collected per second to 25 to support a localization update rate of 1\,Hz. 
The results  in Fig.~\ref{fig:var-refresh-rate} and ~\ref{fig:var-node-count} show that when budget restrictions increase for a given topology size or vice versa, the accuracy degrades by several folds, clearly exposing a trade-off between latency (cost) and accuracy. This can be attributed to the lack of intelligent topology estimation (edge selection) and robust relative localization schemes that are needed to work with limited latency budgets and hence incomplete range estimates respectively -- aspects that have not been addressed thus far.  We now further dissect the specific factors contributing to this performance degradation.

\mypara{Incomplete range estimates:} In practice, the physical communication range between nodes will limit the topology from being complete. 
This is further compounded by the limited number of edges (ranges) that can be estimated due to the latency budget. 
Since an incomplete EDM (i.e. estimated topology is not a clique) can lead to localization inaccuracies, matrix completion methods (e.g., \texttt{SDR}~\cite{keshavan2010matrix}, \texttt{OptSpace}~\cite{alfakih1999solving}) are used to complete the EDM before the nodes can be relatively localized. However, the latter are not designed keeping in mind the geometrical implications relevant to a localization problem. This can lead to large localization errors as shown in Fig.~\ref{fig:var-topologies}, thereby advocating the need for  relative localization schemes that are robust to incomplete range estimates. 

\mypara{Impact of geometry:} The interesting result in Fig.~\ref{fig:var-configurations} further indicates that the specific set of edges selected, albeit incomplete, has a large impact on the localization solution as well. This indicates that the geometry of the topology (particularly its rigidity) associated with the edges measured, has a direct impact on the solution and must be factored into the edge and hence topology selection process.

\mypara{Inaccurate range estimates:} Inaccurate estimates of even a small set of ranges can lead to degraded accuracy for the entire topology. Here, two key environmental factors, namely LOS blockages (due to body, concrete, etc.), and node mobility, can significantly affect the accuracy of the range estimates. Given the limited budget for range estimation, it is clear that when the edges are picked randomly without taking into account their channel or mobility characteristics, the performance degrades quite rapidly even with a small set of affected edges, as seen in Fig.~\ref{fig:var-noisy-link}. 
Thus, characterizing the nature of the edges with respect to their channel and mobility is essential for improved localization accuracy.

\subsection{Design Requirements}
From the above discussions, it is clear that the combination of ``infrastructure-free" and ``node mobility" in practical, dynamic indoor environments, makes the {\em latency-bounded} version of the localization problem, highly challenging. 
In addressing these challenges, two key design requirements emerge for \system,
(a) Support a reasonable number of nodes in a practical deployment setting ($\approx$ few 10s), many of them being mobile ($\leq$ 2 m/s), and (b) Offer a location update rate ($\ge$1\,Hz), tolerable to the underlying location based service that eventually consumes such information. 

\section{{\bf \Large \system}: Design}
    \label{sec:method}
At a high level, \system models the topology among nodes {\em collectively} as a graph of rigid components, and tracks it accurately over time as the topology evolves subject to node mobility and channel conditions. 
Within a limited time (determined by the application's refresh rate), \system's task is to gather as much UWB ranging information (on links) from the network as possible at a master\footnote{One of the nodes doubles up as a master node.} node (using WiFi for control),  so as to accurately estimate the underlying topology (See Fig.~\ref{fig:dynoloc-overview}). 
%
\subsection{DynoLoc in a Nutshell}
\system operates in epochs (rounds), where the locations of all nodes are estimated at the end of each epoch, the duration of which is determined by the application's refresh rate (e.g., 0.5-2 Hz). 
In every epoch, the following sequence of operations is executed.

\mypara{(i) Topology estimation for ranging (Sec \ref{subseq:ranging-tech})}
Given the underlining physical topology (based on connectivity), \system
prioritizes edges which contribute to the resulting topology being {\em maximally rigid}, while avoiding those, whose ranges could be corrupted by multi-path; and it is {\em adaptive} in that it prioritizes edges associated with nodes that have been mobile in the recent past  thereby leading to a good localization accuracy. 
%

\mypara{(ii) Concurrent ranging (Sec \ref{subseq:concur-ranging})} The selected edges are then ranged using a concurrent ranging protocol that amortizes the overhead of ranging from a node across its neighbors, while enabling concurrent ranging in non-interfering neighborhoods, to minimize the overall latency. 

\mypara{(iii) Robust relative localization (Sec \ref{subseq:relative-loc})} After the estimated topology has been ranged, \system's localization algorithm intelligently 
identifies and applies EDM only on sub-graphs of the topology that are rigid and combine them effectively, to deliver both a robust and accurate relative localization. 

\mypara{(iv) Absolute localization (Sec \ref{subseq:abs-loc})} \system  finally transforms the relative localization solution into an absolute one with little additional meta information (contributed by floor plans or a single reference node), while still delivering the desired refresh rate for the solution. 
%
%
%
%
%
%
%
\subsection{Estimation of Ranging Topology}
\label{subseq:ranging-tech}
\system's innovation lies in leveraging the graph theoretic construct of geometric rigidity to help identify the set of edges that would collectively contribute to the accurate localization of the topology as a whole while also adapting itself to track the topology as it evolves with node mobility and channel conditions. \system accomplishes this by  first characterizing the links in the physical connectivity topology, followed by leveraging such a characterization for adaptive link selection.

\mypara{A. \underline{Characterizing the Connectivity Topology:}} 
Every link is characterized based on the mobility of its nodes, the multipath nature of its wireless channel, as well as its contribution to the topology's rigidity. 

\begin{figure}[!htb]
	\centering
	\includegraphics[width = 0.65\linewidth ]{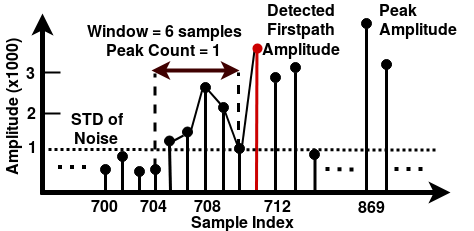}
	\vspace{-0.1in}
	\caption{First Path Detection in UWB receiver}
	\label{fig:l-metric} 
	\vspace{-0.15in}
\end{figure}
\mypara{LOS vs. NLOS:} Every node maintains a list of its neighbors, identified by overhearing their transmissions, whose channel impulse response (CIR) is also collected. A node $i$ can thus directly range with any of its neighbors $j$, whose link quality ($L_{ij}$) is estimated from its corresponding CIR. $L_{ij}$ captures the potential accuracy of ranging on the link based on the certainty of it being a LOS (direct) or NLOS (indirect) path. %
This NLOS probability is computed by as $p_{NLOS} =(f_1\times{}f_2\times{}f_3\times{}f_4)$, where $f_1 =$ avg. peak count before the detected first path (FP) in the preceding window (See Figure \ref{fig:l-metric}), $f_2=$ ratio of std-noise to FP amplitude, $f_3=$ ratio of peak to FP amplitude and $f_4=$ ratio of total received power to FP power. The link quality is the inverse of this probability.
Initially, during the bootstrapping phase, every node sequentially broadcasts a {\em beacon} packet that is heard by its neighbors. Once the system reaches a steady state, the neighborhood list is implicitly maintained by all nodes without the need for additional ranging. This helps realize a physical connectivity graph across the nodes, where every edge is a potential candidate for range estimation, and is weighted by its quality (i.e. certainty for delivering accurate ranges). 

\mypara{Mobility:} In addition, every node $i$ also maintains a mobility metric $M_i$ that capture its location uncertainty since its last localization. This metric increases as a function of  the time-since-localization (TsL) and is computed using the node's acceleration, $a_i$ (\sfrtxt{\uline{obtained from its IMU}}). In particular, for every IMU read (indexed by $k$), $M_i(k) \leftarrow M_i(k-1)+v_i(k)\cdot \Delta t$, where node velocity $v_i(k) \leftarrow v_{i}(k-1) + a_i\cdot \Delta t$, and $\Delta t$ is the elapsed time since the last IMU read.  
$M_{i}$ and $v_i$ are reset to zero, whenever the node is localized. When $a_i$ is zero (static nodes), we assign an exponential function to $M$ as follows: $M_i \leftarrow (e^{TsL} -1 )$. This allows the node to be prioritized for ranging, even if it is static, but sufficient time has elapsed since its last localization. 

\mypara{Geometric Rigidity:} Recall that a rigid graph admits a unique relative localization solution. Since the connectivity  topology of nodes might not be a rigid graph in practice,  \system aims to select edges from this underlying connectivity that ensures {\em maximal} rigidity to the resulting node topology.
It does so by identifying maximal rigid sub-graphs from the physical connectivity graph, by leveraging the construct of $k$-core sub-graphs that are used to ensure graph rigidity~\cite{eren2004rigidity}. In a $k$-core sub-graph, every vertex has a degree of at least $k$. It is known that a $k$-core sub-graph is rigid in $k-1$ dimensional space~\cite{dokmanic2015euclidean}. Hence, for rigidity in 2D, we seek to obtain 3-core sub-graphs\footnote{This allows \system to also be extended for 3D localization, where 4-core sub-graphs will be leveraged instead.}. Note that a 2-core sub-graph will not be rigid in 2D and will admit multiple localization solutions. 

\begin{figure}[htb!]
	\centering
	\includegraphics[width = 1.01\linewidth ]{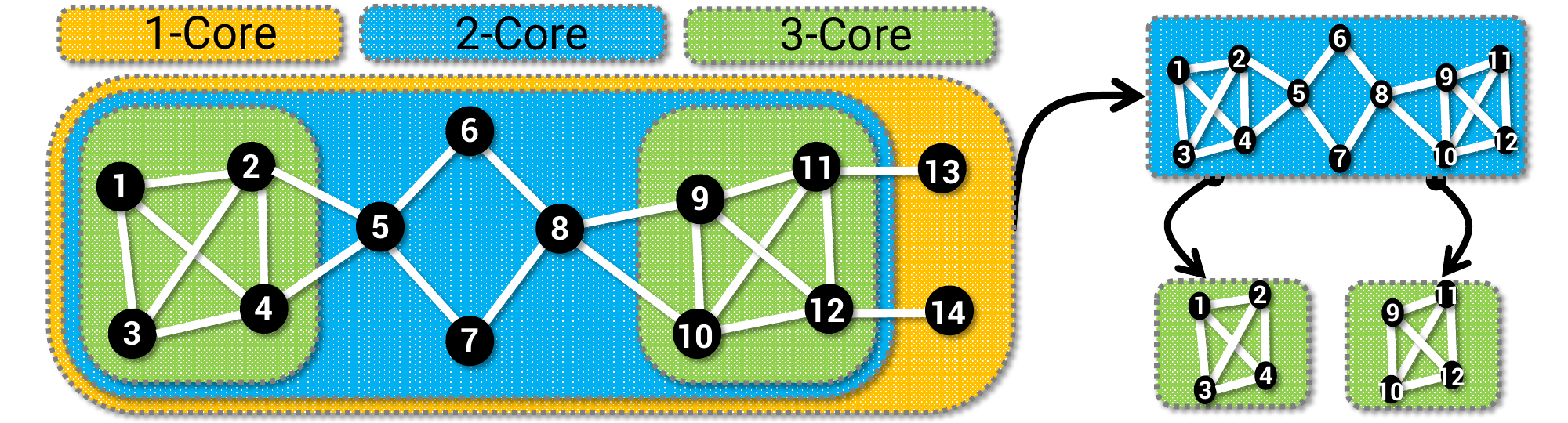}
	\vspace{-0.15in}
	\caption{Example of k-core Decomposition}
	\label{fig:k-core-example}
	\vspace{-0.15in}
\end{figure}

\system identifies the maximal 3-core sub-graphs by starting with the connectivity graph and partitioning it into $k$-core subgraphs for k = 1,2 and 3 sequentially. It starts with identifying 1-core nodes (one by one) that have a degree of 1 and removes them and their incident edges iteratively till no more 1-core nodes can be found. Then, it repeats the process for 2-core nodes with degree 2. After the removal of 1-core and 2-core nodes, we are left with maximal 3-core sub-graphs (as shown in the example in Fig.~\ref{fig:k-core-example}) that are rigid. 

\mypara{B. \underline{Estimating the Ranging Topology:}} 
We now describe  
\system's algorithm for edge and hence topology selection that will be used for ranging.
At a high level, \system aims to devote its ranging resources to links, whose ranges are outdated  (due to mobility), followed by those that contribute the most to the topology's rigidity, while also avoiding those with potentially inaccurate ranges (due to NLOS).
Specifically, at every iteration, \system picks the node (say $i$) with the highest mobility $M$ (location uncertainty) metric. If $i$ is part of the 3-core, and has more than three edges, then three of its edges with the  highest $L$ metric (range accuracy) are selected. Otherwise, its incident edges ($\le$ 3) are directly selected. When multiple nodes have the same $M$ metric, the node selection is done based on the $k$-core metric, with nodes belonging to a higher core (ties broken with higher node degree) prioritized over those belonging to a lower core. The process repeats until the ranging budget is exhausted by the edges selected for ranging. Initially, when the system is bootstrapped and no information on node mobility is available, all nodes are assumed to have outdated $M$ metric and edge selection is done primarily based on their contribution to rigidity and their LOS nature. 
The complete  DynoLoc method is presented in Algorithm \ref{algo:dynoloc}.

\begin{algorithm}[htb!]
	\caption{DynoLoc Algorithm}
	\label{algo:dynoloc}
	\begin{algorithmic}[1]
		\scriptsize
		\STATE Make every node send hello frame 
		\STATE Collect and initialize link quality metric $L$
		\STATE Run core-decomposition based on $L$
		\STATE Build initial rigid graph $G_R$
		\STATE Initialize mobilty metric $M$
		\WHILE{\textbf{True}}
		\STATE Choose node $i$ with the highest metric $M(i)$
		\STATE Remove $i$ (with its edges) from $G_R$ or $V^{'}$
		\IF{$i$ has $\geq$ 3 edges with $G_R$}
		\STATE Select 3 edges for $i$ with max. $L$
		\STATE Add $i$ back to $G_R$ with selected edges
		\ELSE
		\STATE Put $i$ in the excluded node list $V^{'}$ 
		\STATE Select $i$'s edge(s)
		\ENDIF
		\STATE Range for the above-selected edges [\textbf{\S 3.3}]
		\STATE  Set $M(i) \leftarrow 0$
		\STATE Update $M$ \& $L$ metrics from collected data
		\STATE Compute core-decomposition
		\STATE Remove  nodes which are not in 3-core, from $G_R$
		\IF{Refresh interval elapsed}
		\STATE Complete EDM for $G_R$ [\textbf{\S 3.4.A}]
		\STATE Determine relative locations [\textbf{\S 3.4.B}]
		\STATE Determine \& output absolute locations [\textbf{\S 3.5}]
		\ENDIF
		\ENDWHILE
	\end{algorithmic}
\end{algorithm}
\vspace{-15pt}
\subsection{Aggregated and Concurrent Ranging}
\label{subseq:concur-ranging}
\system  optimizes its ranging protocol using two key mechanisms as follows.  
 
%
%

\noindent\textbf{Aggregated Ranging:} 
\sfrtxt{ \system{} employs two-way ranging (TWR) mechanism \cite{sahinoglu2006ranging} further optimized for tight TDMA control. See Appendix A for details.}

\mypara{Concurrent Ranging:} 
\sfrtxt{
We use concurrent ranging to increase ranging budget and hence support more nodes for a given refresh rate. Discussion about concurrent ranging is presented in Appendix B.
}
%
%
\subsection{Robust Relative Localization}
\label{subseq:relative-loc}
\system employs the latest (epoch) ranges collected from the ranging topology, along with other ranges from prior epochs that are accurate and not outdated, to determine the relative localization of nodes.
Existing solutions require a complete $m\times m$ EDM matrix of ranges for a $m$ node topology to determine the relative localization (embedding) of nodes. 
Matrix completion techniques (e.g. OptSpace~\cite{keshavan2010matrix}, SDR~\cite{alfakih1999solving}) are used to complete the missing EDM entries in practice, but deliver poor accuracies, owing to the lack of (i) rigidity over the entire topology, and (ii) incorporation of geometric structure. 
\system innovates in both these aspects to deliver a robust relative localization solution as follows. 


\mypara{A. \underline{Estimating Missing Ranges:}} 
\system performs EDM completion only on the rigid sub-graphs (3-core sub-graphs) of the topology, individually. 
The rigidity of the sub-graph (say, with $n$ nodes) enables a more accurate EDM completion of $n\times n$.
Further, it leverages the geometric topology of the nodes to complete the missing EDM entries as follows (Algorithm \ref{algo:edm-completion}).
The algorithm starts by initializing the EDM with the measured ranges and then fills in the missing ranges through sequential multilateration. In the process, it also computes the relative locations of all the nodes. It maintains two copies of the EDM, one based on the inter-node ranges obtained  solely from the computed relative node locations, and the other including the actual measured ranges, where available. It then iteratively perturbs or updates the relative location of each node to minimize the gap between these two EDMs, and further updates the EDM from the subsequent node locations. Thus, the error in the estimation of missing ranges is minimized. \sfrtxt{See Appendix C for EDM completion algorithm.}
\mypara{B. \underline{Relative Localization:}}
Having identified the rigid and non-rigid components of the topology, \system first solves for the relative localization of nodes in the rigid sub-graphs individually, followed by those in the non-rigid components. 
%
\\\noindent\textit{\underline{Rigid Component:}} 
Having completed the EDM for the rigid component, \system employs the Multi-dimensional Scaling (MDS) solver~\cite{costa2006distributed} to find the node embedding (relative localization) solution. Note that while sequential multilateration is appropriate for EDM completion, it is not employed for the eventual localization itself, as the location error tends to increase rapidly for the nodes being mulitlaterated later. In contrast, with MDS solvers, the order of the multilateration does not matter -- with the multilateration error being part of the objective function, it is equally distributed across all steps of the multilateration process. 
The MDS problem can be defined as: Given a squared EDM $\mathbf{D^2}$, find 
the corresponding relative location matrix $\mathbf{X}$. In matrix notation, this is equivalent to solving for $\mathbf{B}=\mathbf{XX^T}$ where $\mathbf{X^T}$ is the transpose of matrix $\mathbf{X}$. \system employs the Classical MDS (CMDS) solver \cite{costa2006distributed, dokmanic2015euclidean}, which works as follows:
%
%
\begin{packeditemize}
	\item  Given square-distance matrix $\mathbf{D^2}$, compute Gram matrix $\mathbf{B}$ of $\mathbf{X}$ as $\mathbf{B}=-\frac{1}{2}\mathbf{JD^2J}$, where $\mathbf{J}=\mathbf{I}-\frac{1}{n}\mathbf{1\,1^T}$
	\item Find the Eigenvalue decomposition of $\mathbf{B}=\mathbf{Q}\,\Lambda\,\mathbf{Q^T}$, 
	\item Finally, compute $\mathbf{X}=\mathbf{Q}\sqrt{\Lambda} $ for Eigenvector matrix $\mathbf{Q}$ and diagonal Eigenvalue matrix $\Lambda$.
\end{packeditemize}
Due to the ``centering" assumption, each column of solution $\mathbf{X}$ sums to zero i.e, the origin of $\mathbf{X}$ coincides with the centroid of the $n$ locations. 
Note that CMDS minimizes the loss function (also called \textit{strain}) $L( \mathbf{X} ) = ||\mathbf{XX^T} - \mathbf{B}||=||\mathbf{XX^T} - \frac{1}{2}\mathbf{JD^2J}||$. In doing so, it updates the  range entries only for the missing edges but leaves the edges with measured ranges untouched. The CMDS can result in local minima, which are more likely to occur, when the dimension is low (2  as in our case) ~\cite{costa2006distributed, dokmanic2015euclidean}. 
Hence, \system cross-validates the locations using heading data retrieved from the IMUs of nodes and also employs a smoothing filter on the successive node locations over a small window. 

\noindent\textit{\underline{Non-Rigid Component:}}  
Having solved for the relative locations of the 3-core nodes, \system now considers the nodes in the non-rigid components, namely the 2-core and 1-core nodes. By definition, every 2-core node has two edges to the rigid component(s). It employs these two ranges to find the relative location of the 2-core node. However, with just two ranges, there are two possible solutions. \system eliminates one of these solutions easily by leveraging the  link quality of edges  in the neighborhood of the solution. If the solution has a neighboring rigid node (other than the rigid nodes of its two edges), this would contradict its 2-core status (as another edge to a rigid component exists) and the solution can hence be eliminated. 
Once the 2-core nodes are also solved, the remaining 1-core nodes are determined using the  angle that the corresponding edge (to the rigid component) makes with the North-South axis of the earth using the IMU data. This computation is done as part of the absolute localization procedure discussed in \S{}3.5.
\\
\sfrtxt{
\noindent{\underline{Continuous Link Info. Update:}} \system{} nodes continuously overhear all wireless frames and decode the sources and update the link quality metrics as per Line 18 of Algorithm \ref{algo:dynoloc}. If no frame is received or overheard within a threshold period (30 seconds for our operation), then the link metric value for the source node is set to zero, which means reachability of the node. The controller forces a node which has not sent any frame within the threshold period, to range with neighbors, thus facilitating continuous update of physical connectivity. 
}
%
\vspace{-1em}
\subsection{Absolute Localization}
\label{subseq:abs-loc}
%
\system needs to transform the localization solution from a  relative coordinate system to a target absolute coordinate system. This is achieved by translating, rotating, and flipping the rigid graph derived in the previous step with the help of IMUs (on the nodes) or floormap information \cite{xie2016xd}. 
\sfrtxt{Details is presented in Appendix D.}
\section{Implementation and Evaluation}
%
\vspace{-0.05 in}
\subsection{System Implementation and Prototype}
\system system consists of a set of {\em UWB tags}, interfaced with an {\em embedded computer} (e.g., a smartphone or a Raspberry-Pi) and a {\em central controller} that orchestrates range measurements and runs the localization algorithms. Both, the controller and the embedded computer are connected to a local WiFi network for exchanging control information and application data (e.g., sensor readings, video feeds).

\begin{figure}[!htb]
	\centering
	\includegraphics[width = 1\linewidth ]{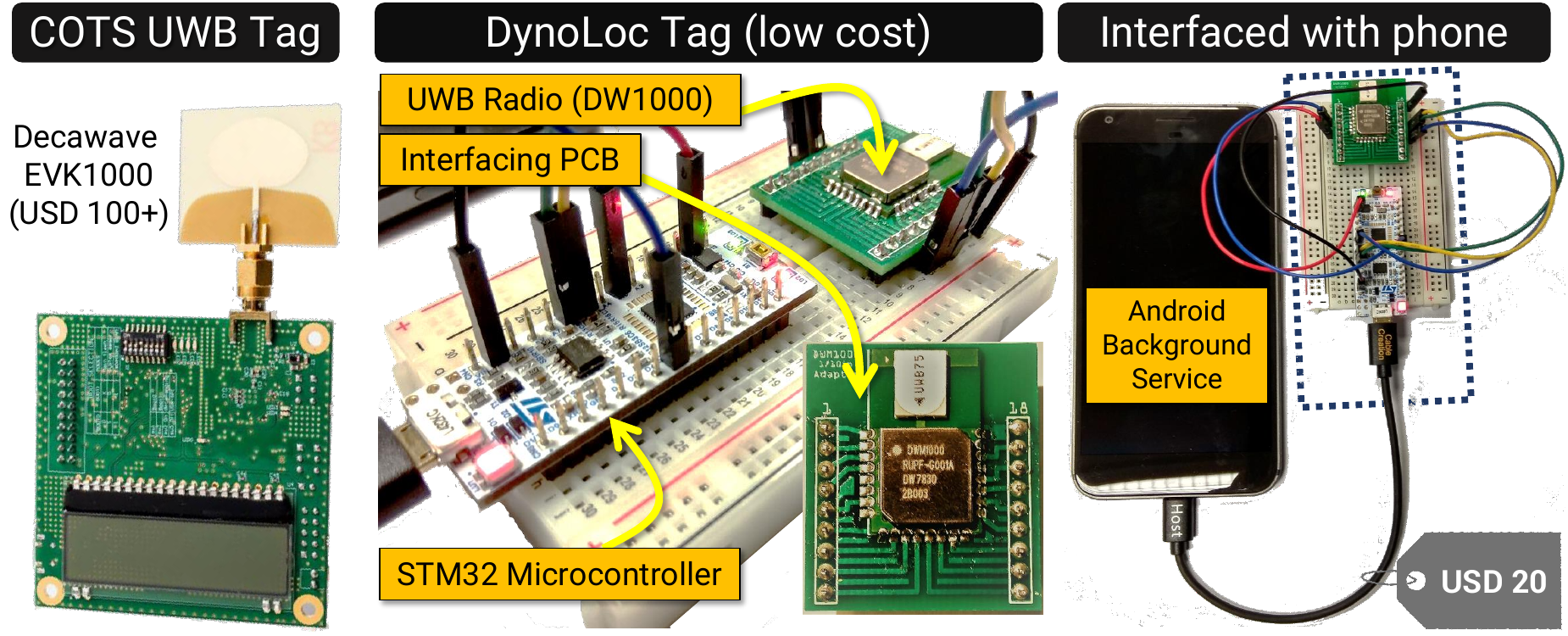}
	\vspace{-0.05in}
	\caption{Prototype of DynoLoc node}
	\label{fig:system} 
	\vspace{-0.05in}
\end{figure}

\begin{figure*}[!htb]
	\centering
	\subfloat[]
	{
		\includegraphics[width=0.25\linewidth]{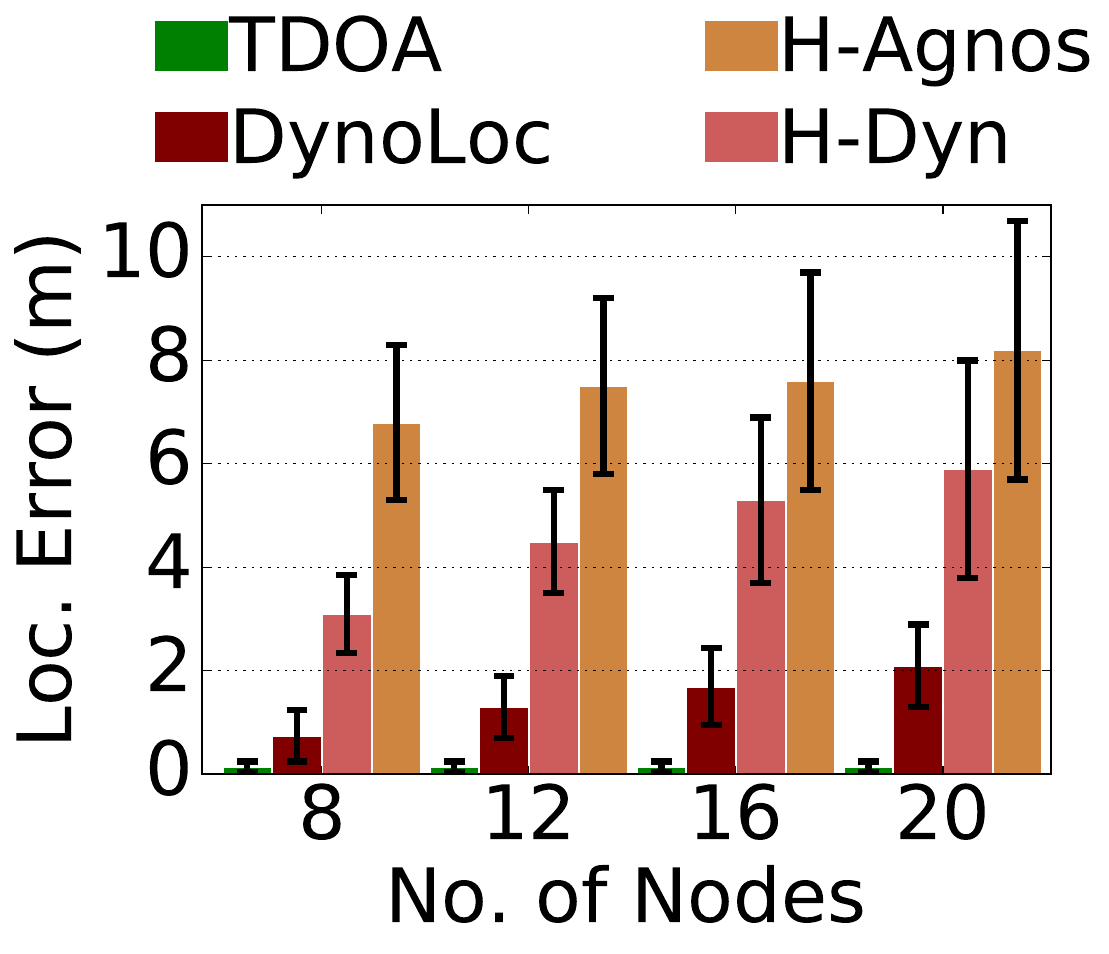}
		\label{fig:base-var-node}
	}
	\subfloat[]
	{
		\includegraphics[width=0.25\linewidth]{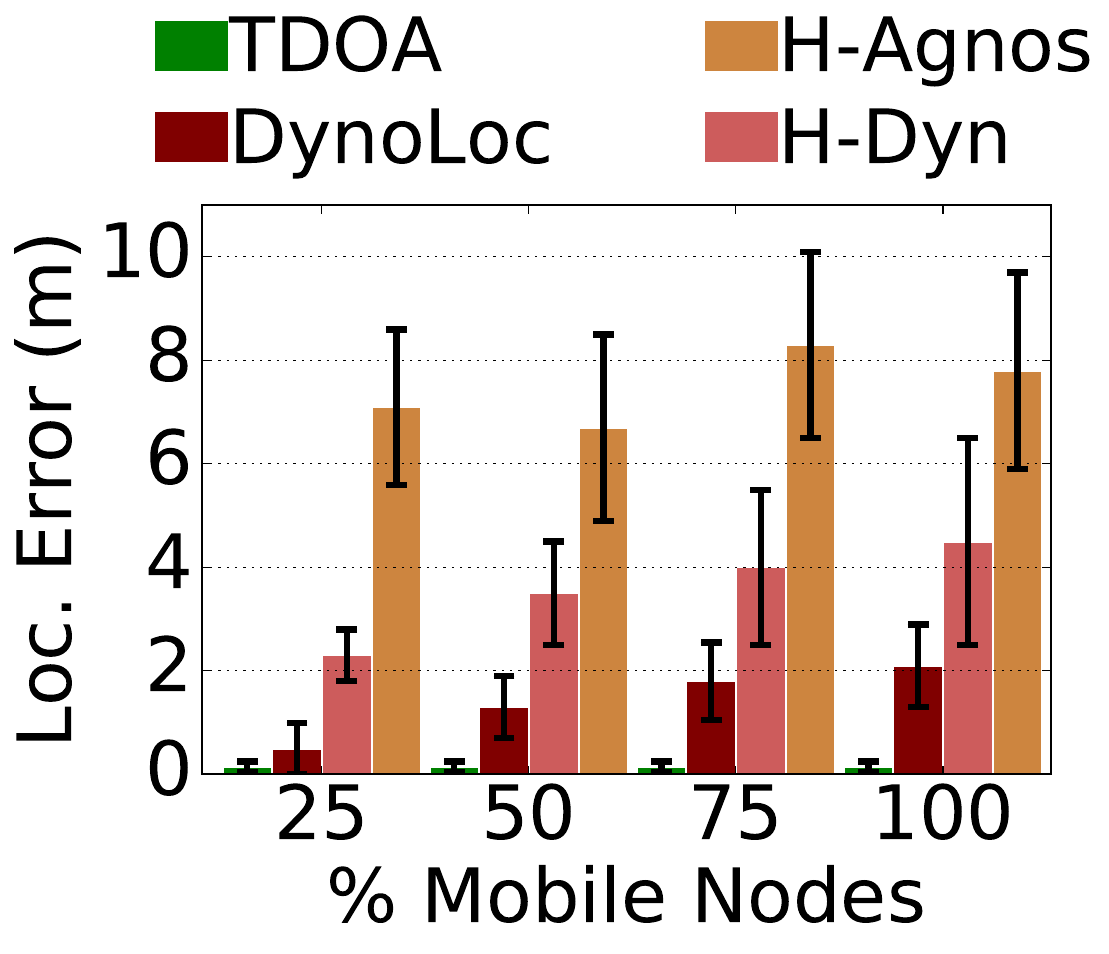}
		\label{fig:base-var-percent-mobile}
	}
	\subfloat[]
	{
		\includegraphics[width=0.25\linewidth]{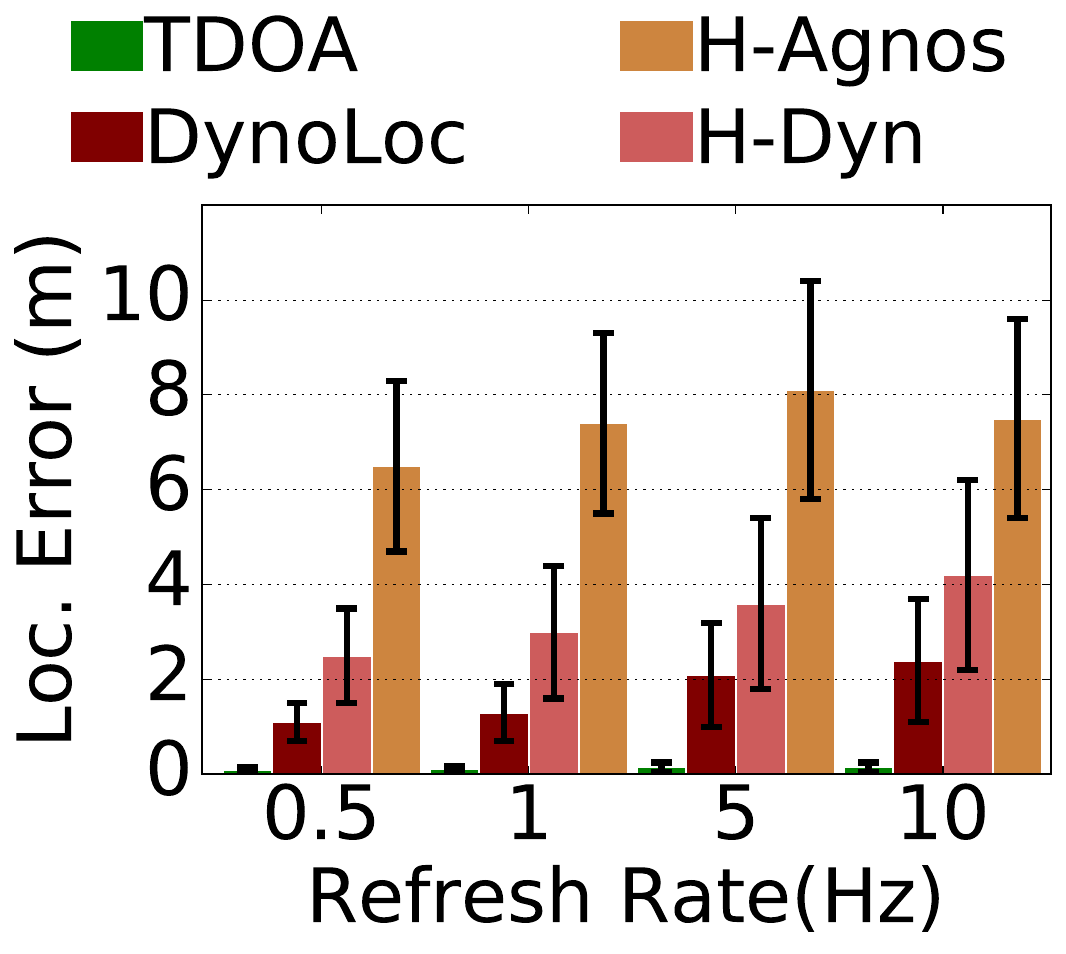}
		\label{fig:base-var-refresh-rate}
	}
	\subfloat[]
	{
		\includegraphics[width=0.25\linewidth]{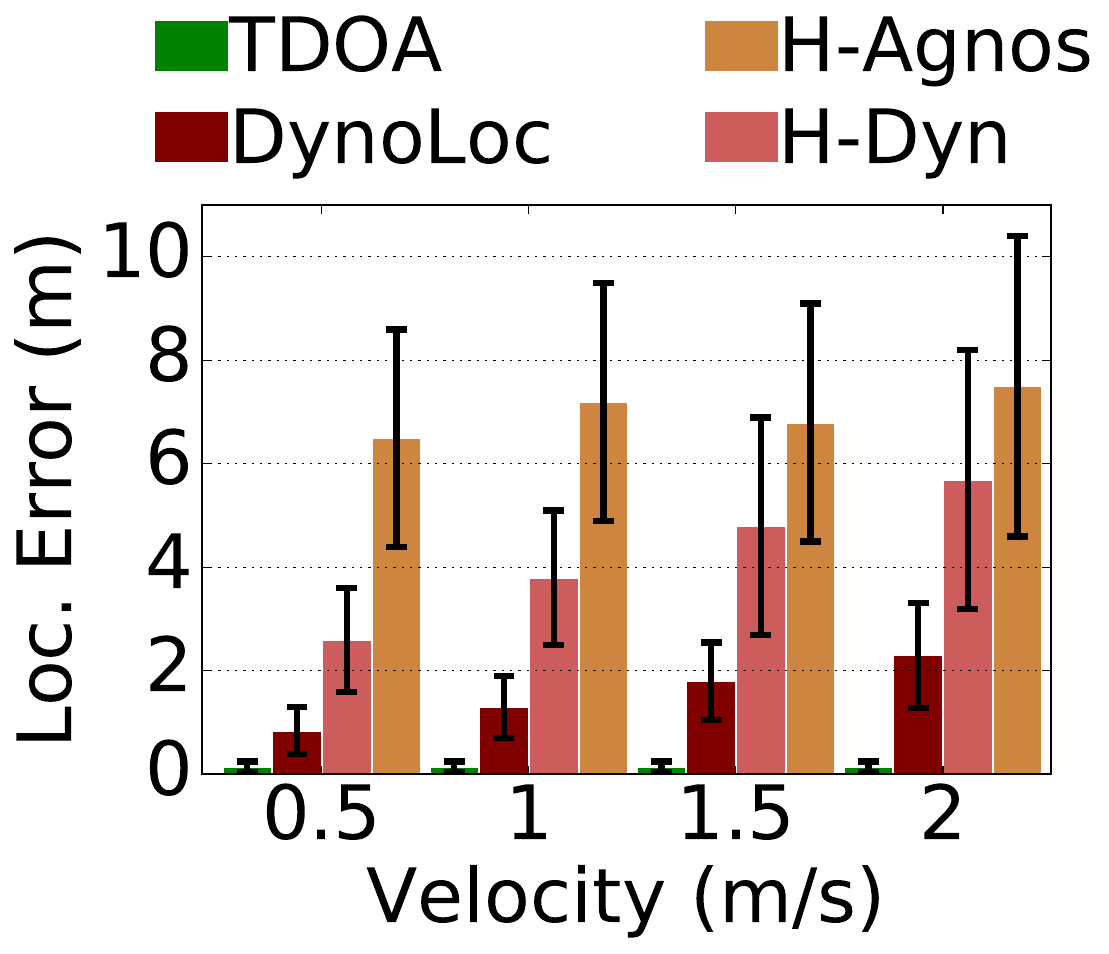}
		\label{fig:base-var-velocity}
	}
	\vspace{-2 pt}
	\caption{Comparison of localization error for various method of localization, default values: no. of nodes = 12, velocity = 1m/s, refresh rate = 1Hz, fraction of mobile nodes =50\%}
	\vspace{-5 pt}

	\label{fig:base-var}
\end{figure*}

\noindent\textbf{$\blacksquare$ \system UWB Tag:} The tag consists of a Decawave DW1000 \cite{decawave} ultra-wideband radio (costs $\approx$10\$) that houses an extremely precise picosecond crystal (for TOF calculation), which can achieve a distance resolution as high as 2.2\,mm~\cite{decawaveUserManual}. We interface the DW1000's \texttt{SPI} pins (serial clock, master output, master input and slave select), \texttt{V$_{cc}$} and \texttt{GND} pins to a low power ARM Cortex-M based microcontroller unit STM32 NUCLEO-F042K6 (costs $\approx$10\$), where the latter acts as the \texttt{SPI} master. \system's optimized ranging protocol is implemented in about 4000 lines of \texttt{C} code and runs on the STM32 microcontroller. Additionally, the STM32 sends and receives specific ranging instructions or estimated ranges through its serial port from an external host device. While most commercially available UWB tags cost somewhere between \$100--200, \system tags show that it is feasible to keep the cost to under 20\$ (10-20\% of the COTS tags) without compromising any feature. Our tags can use the UWB permissible channels spanning from 3--7\,GHz (with 500\,MHz bandwidth). In most of the experiments, we use 3.5\,GHz as the center frequency for improved range. 

\noindent\textbf{$\blacksquare$ Tag Host:} An embedded computer (e.g., a smartphone) acts as a USB host for the \system Tag. It sends specific ranging instructions to the tag (e.g., range with 5 specific neighboring nodes) and receives measured ranges. The tag host also keeps track of the node's mobility from the inertial sensors as well as the link quality information as obtained from the tag ($M$ and $L$ metrics). The device driver for our tag is implemented on the Android platform and runs as a background service intercepting commands from the controller and passing it on to the tag and vice versa. The tag host also houses a pressure sensor to identify the vertical elevation, i.e., floor number which is useful in a multi-storeyed deployment scenario.

\noindent\textbf{$\blacksquare$ Controller:} The controller is in charge of the overall topology estimation, gathering ranges from individual tags and running the localization engine. Depending on the application (see more in \S~\ref{sec:applications}), the controller sends the information back to the individual nodes, displays them locally in a dashboard or offload it to a cloud service for remote visualization or decision making. The controller logic is implemented in about 1000 lines of \texttt{Python} code.
\subsection{System Evaluation}
\system is evaluated comprehensively spanning {\em in-the-wild} deployments to controlled experiments in realistic indoor settings (supplemented by simulations only for larger topologies of over 16 nodes). 
In the following, we describe our methodology followed by some key performance results.
\vspace{-0.25 cm}

\subsubsection{In-the-wild Deployment}
\label{subsec:in-the-wild}
\system{} has been deployed and tested in a real firefighter's drill. A total of ten firefighters, each carrying a \system{} tag individually, enter the test building (2 floors, each $\approx$50\,m$\times$100\,m) emulating a severe fire incident. A pressure-sensor+UWB based mechanism 
\sfrtxt{(See Appendix E)}
 for floor identification at each node locally, was added to \system for this drill. The fire chief, stationed outside the building, tracks every move of the personnel crawling through the dark and smoky passages through \system's real-time dashboard and instructs them accordingly through a walkie-talkie. In one specific incident, a firefighter who was lost and separated from his colleagues issued an SOS call. \sfrtxt{The chief knowing his location from the breadcrumb feature of \system{} app (See Fig.
\ref{fig:vertical-detection}(b) 
in Appendix E) was able to intervene and immediately assist by redirecting his crew accurately towards the lost firefighter.} 
We learned (from the fire chief) \system's true value in such challenging scenarios, which are common and often lead to firefighter fatalities. 
A snapshot of the drill is shown in Fig.~\ref{fig:dynoloc-overview}.
%
\subsubsection{Controlled Experiment}
Here, we describe results derived from known path and speed of the mobile nodes.
\\
\textbf{Testbed:} The testbed consists of eight pre-planned navigable routes (marked with adhesive tapes on the floor) in an indoor area of about 50\,m$\times$40\,m. The collective length of the routes is $\approx$500\,meters The routes encompass various types of indoor areas: open hallways, corridors, meeting rooms, lab spaces and so on, such that we have a fair representation of typical indoor settings (both LOS and NLOS). In addition to the \system tag (+ smartphone host), each volunteer carries another UWB tag operating at a different frequency (6.5\,GHz) for ground truth collection. 
%
%
%
\\
\noindent{}
\textbf{Baselines:} The ground-truth is collected using a system of densely placed, synchronized, static anchor nodes (by multilateration using TDOA information), deployed throughout the building floor, which gives a localization accuracy in the order of 10--20\,cm, thereby serving as a lower bound on performance for infrastructure-free solutions. 
\sfrtxt{Further, the volunteers carrying the \system{} nodes followed paths of known shapes (see Fig.\ref{fig:exp-paths} in Appendix D). As such, any deviation from the target shape was considered as the error in TDOA itself, which is $\leq$10cm.}
%
We also consider two heuristics that are subsets of \system, namely \texttt{H-Agnos} and \texttt{H-Dyn}. 
\texttt{H-Agnos} employs \system's relative localization component but adopts a naive edge selection approach (ranges every pair of possible edges through round-robin) that does not account for node dynamics and link quality. In contrast,  \texttt{H-Dyn}'s edge selection accounts for node mobility by ranging on edges incident with the most dynamic nodes, but disregards the geometric rigidity requirement in its relative localization. 
\sfrtxt{
Note that, SnapLoc\cite{grobetawindhager2019snaploc} and TrackIO\cite{dhekne2019trackio} are two recent works that uses UWB-range based indoor localization. However, SnapLoc is fixed anchor-based and only considers a small indoor space with constant LOS, whereas TrackIO is infrastructure-full (drone-based) yet is outperformed by \system{} for similar node count and mobility.
}
%
\\\noindent\textbf{Overall Localization Performance:} Fig.~\ref{fig:base-var} highlights the overall performance of \system{} as a function of  various factors, namely number of nodes, fraction of mobile nodes, their velocity and the targeted location update rate. \system scales well for a reasonable number of nodes\,(Fig.~\ref{fig:base-var-node}) that is practical in most real life contexts. Even in challenging scenarios, where all the nodes are mobile (at 1 m/s), \system provides an average localization error of under 2 meters for a 1 Hz update rate, 
%
while \texttt{H-Agnos} and \texttt{H-Dyn} incur an error of 6--7 meters and 5--6 meters respectively (Fig.~\ref{fig:base-var-percent-mobile}). Lack of \system's robust relative localization component (included in \texttt{H-Agnos} and \texttt{H-Dyn}), will only lead to further degradation. 
While \texttt{H-Agnos} accounts for underlying graph rigidity, it is devoid of the notion of node mobility and link-quality and hence, renders poor  accuracy; whereas \texttt{H-Dyn} takes into consideration the node mobility, but renders poor accuracy due to lack of enforcing rigidity requirement.
%
Also note that, even for a more demanding location update rate of 2\,Hz, \system maintains a sub-meter localization accuracy (Fig.~\ref{fig:base-var-refresh-rate}), even when half the nodes are mobile. \system's ranging algorithm being mobility-aware, adaptively expends the available time resources in collecting the most critical ranges. This allows it to deliver under 2m accuracy even when 4 times the update rate is desired. 
\\\noindent\textbf{Benefits of \system's Design:}
\system's design components are benchmarked in isolation as follows.

\noindent\underline{\textit{Adaptive Ranging:} }
 Fig.~\ref{fig:base-var} clearly shows the significant merits and usefulness of \system's adaptive ranging (edge selection) mechanism (compared to random selection in Rand). We now explore the merits of its mobility and link quality metrics as part of its topology estimation. 

\begin{figure}[!htb]
	\centering
	\subfloat[]
	{
		\includegraphics[width=0.4\linewidth]{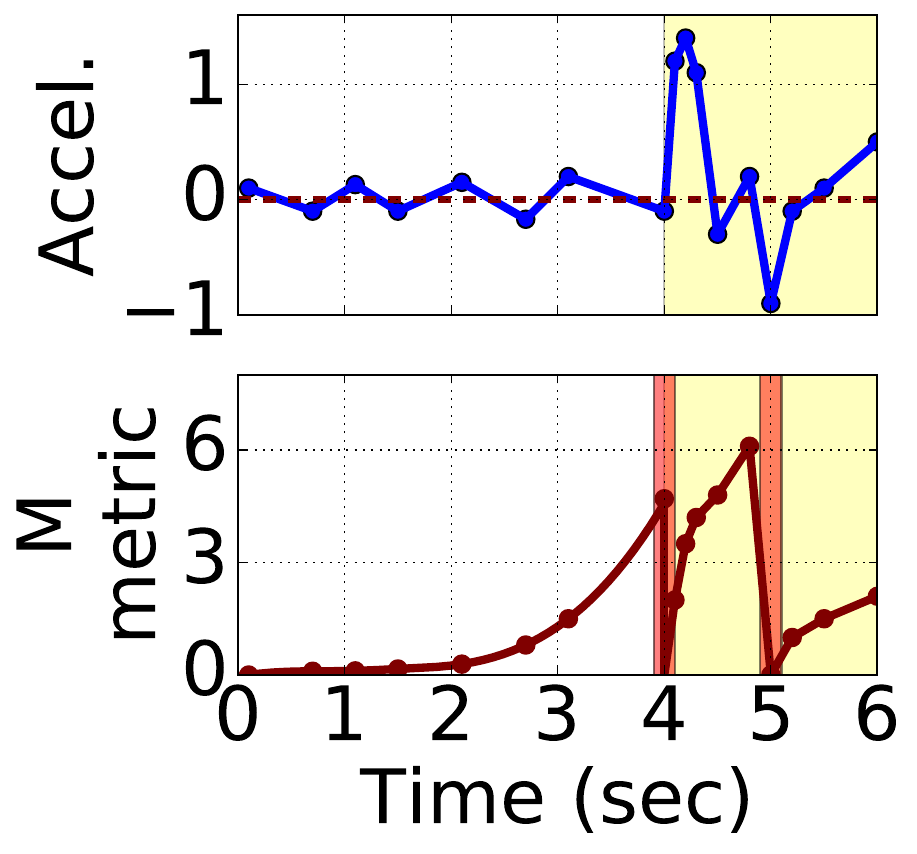}
		\label{fig:mobility-accuracy}
	}
	\subfloat[]
	{
		\includegraphics[width=0.47\linewidth, height=92pt]{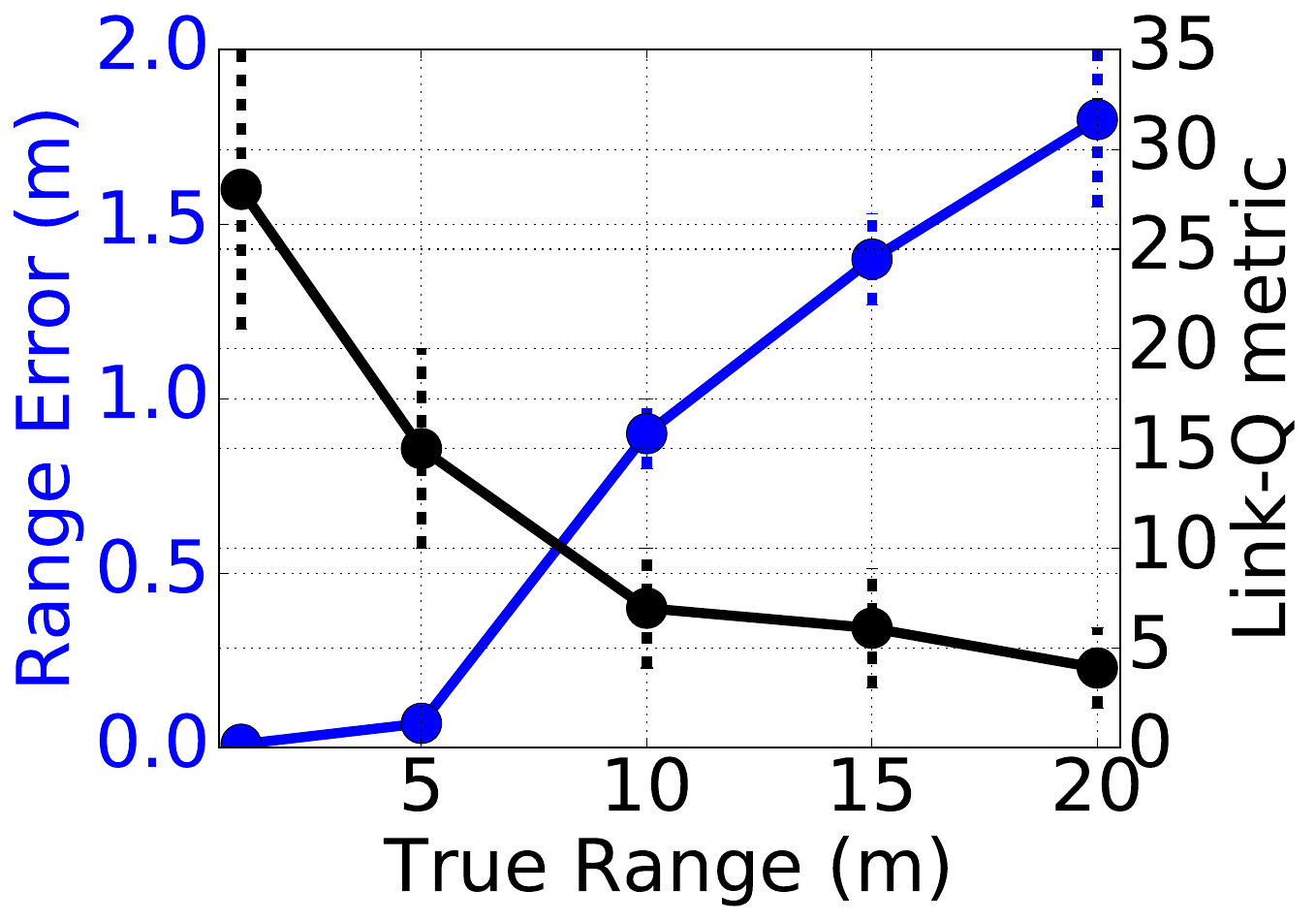}
		\label{fig:linkq-perf}
	}
	\caption{(a) Consistency of Mobility Metric (b) Correlation of Link Quality metric with ranging error}
	\vspace{-12 pt}
	\label{fig:mobility-heading-accuracy}
\end{figure}
\begin{figure}[!htb]
	\centering
	\subfloat[]
	{
		\includegraphics[width=0.5\linewidth]{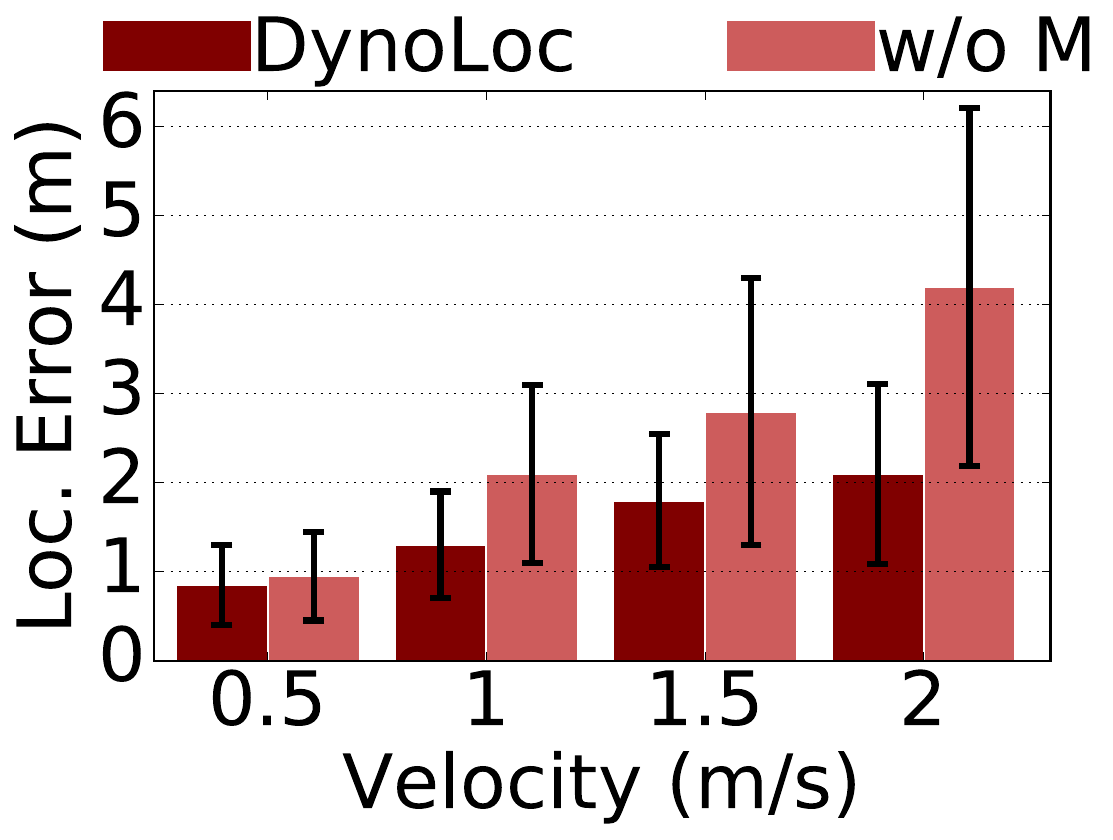}
		\label{fig:effective-m}
	}
	\subfloat[]
	{
		\includegraphics[width=0.5\linewidth]{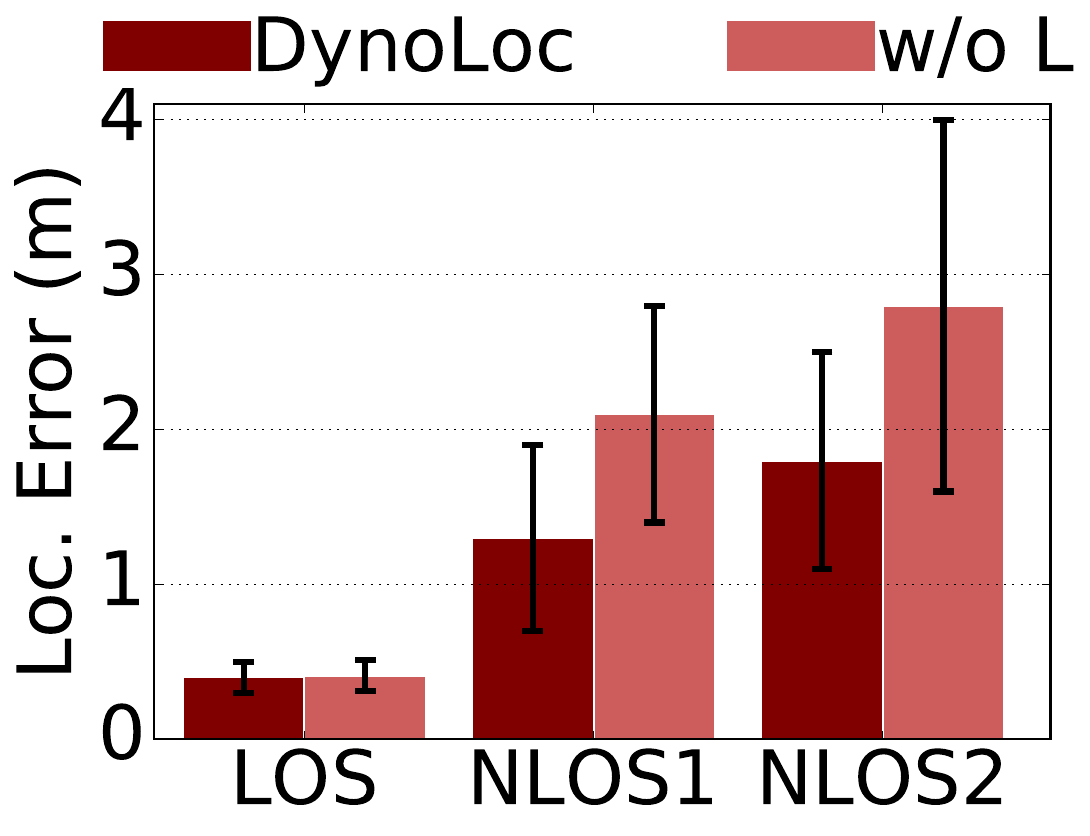}
		\label{fig:effective-l}
	}
	\caption{Comparision of DynoLoc without Mobility\,(M) and Link Quality\,(L) metrics}
	\label{fig:dyno-metric}
\end{figure}

\noindent\textit{$\bullet$ Mobility Metric:} 
Fig.~\ref{fig:effective-m} demonstrates that performing the edge selection (purely based on rigidity constraints) without considering mobility metric ($M$) results in a sub-optimal localization accuracy.  Additionally, Fig.~\ref{fig:mobility-accuracy} shows the reactive nature of the $M$ metric in tracking node mobility through its acceleration.

\noindent\textit{$\bullet$ Link Quality Metric:} Similarly, Fig.~\ref{fig:effective-l} shows the impact of link quality metric ($L$) on edge selection. $L$ acts as a classifier for NLOS versus LOS ranges. Particularly in NLOS scenarios, $L$ plays a critical role in selecting non-noisy edges, thereby leading to a better localization accuracy. Fig.~\ref{fig:effective-l} indicates that introducing the L metric improves accuracy by 30 -- 40\% in two different NLOS scenarios (meeting rooms/office space and lab spaces denoted by NLOS1 and NLOS2 respectively).

\begin{figure}[!htb]
\centering
\subfloat[]
{
	\includegraphics[width=0.4\linewidth]{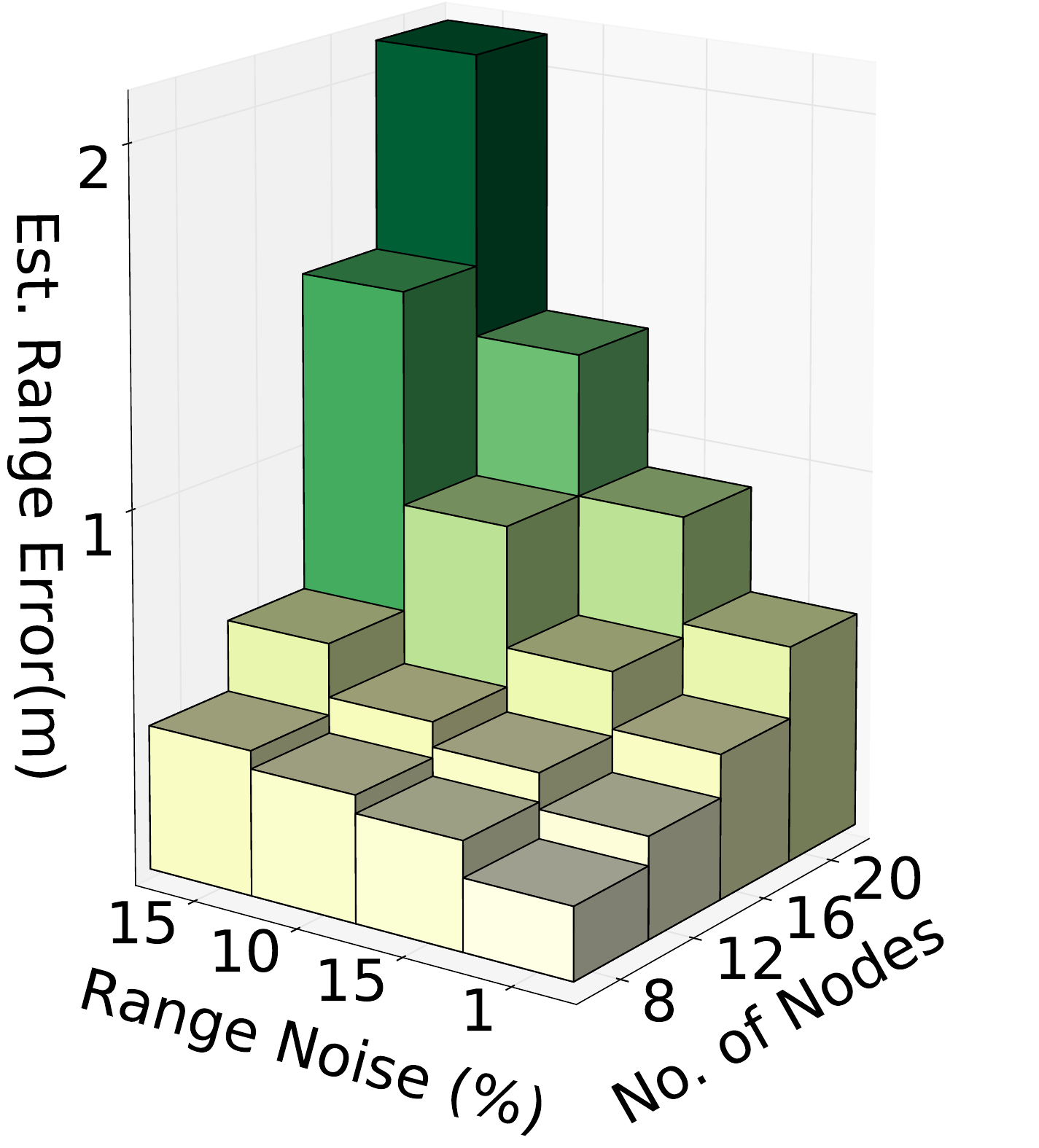}
	\label{fig:edm-completion-error}
}
\subfloat[]
{
	\includegraphics[width=0.4\linewidth]{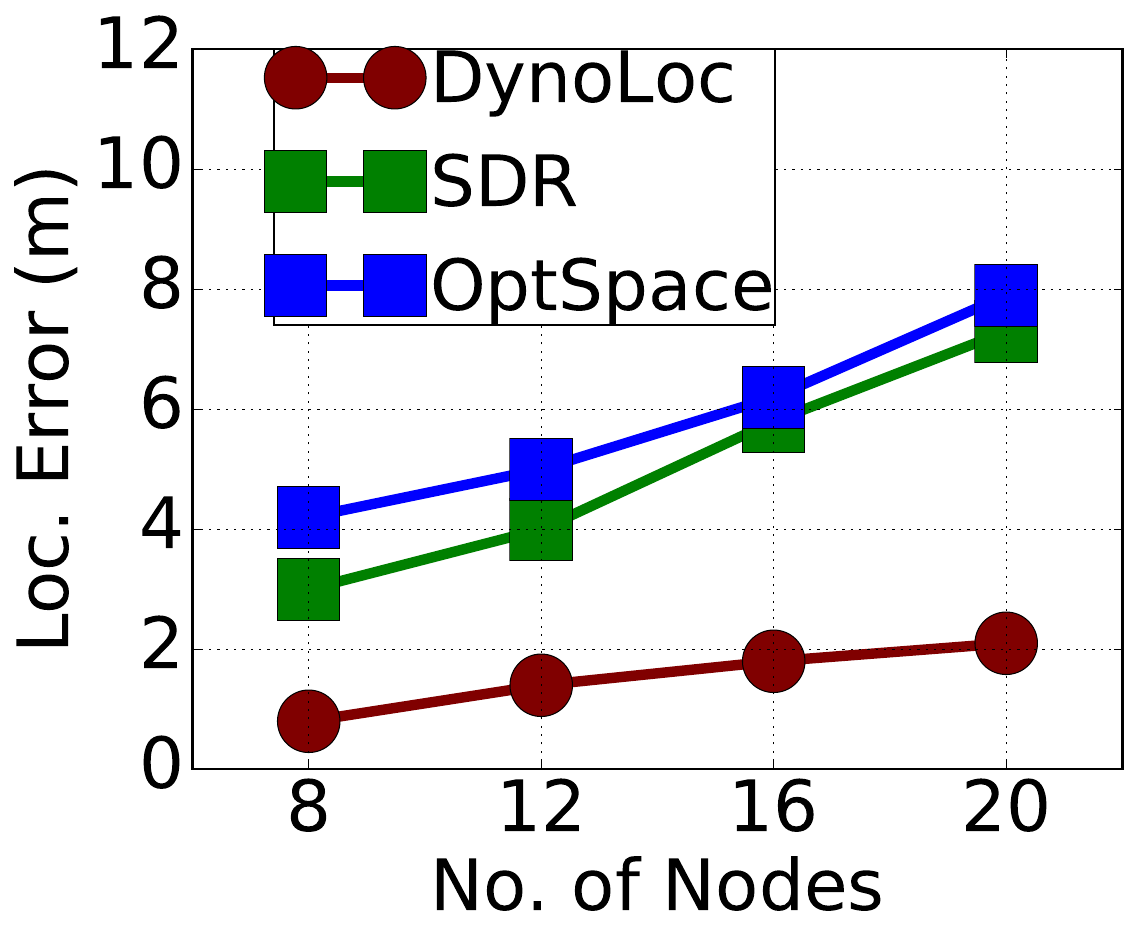}
	\label{fig:edm-vs-multilat}
}
\caption{(a) EDM completion Error, (b) Relative localization error}
\label{fig:edm-related}
\end{figure}

\noindent\underline{\textit{Robust Relative Localization:}} 
Fig.~\ref{fig:edm-completion-error} shows that \system{}'s EDM-completion approach is able to bound the range estimation (for missing entries) to within 2m in most cases. This can be attributed to its approach of targeting the rigid sub-graphs individually, while using a sequential multilateration approach that exploits the node geometry for estimating the missing ranges. Its rigid sub-graph based relative localization further contributes to a much improved accuracy over existing approaches as seen in Fig.~\ref{fig:edm-vs-multilat}.
%
%
%

\begin{figure}[!htb]
\centering
\subfloat[]
{
	\includegraphics[width=0.3\linewidth]{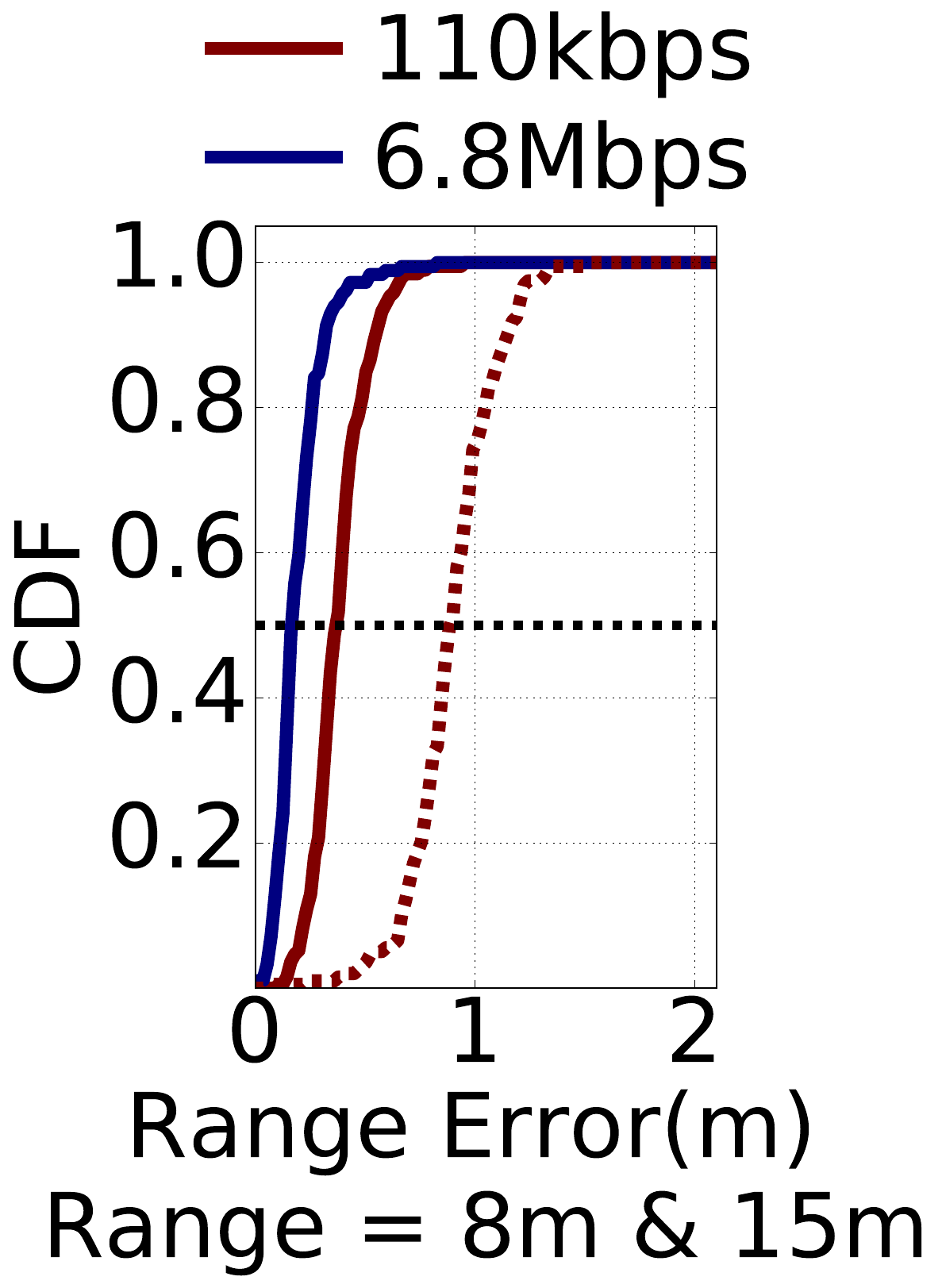}
	\label{fig:diff-data-rate-ranging-cdf}
}
\subfloat[]
{
	\includegraphics[width=0.35\linewidth, height =  97pt]{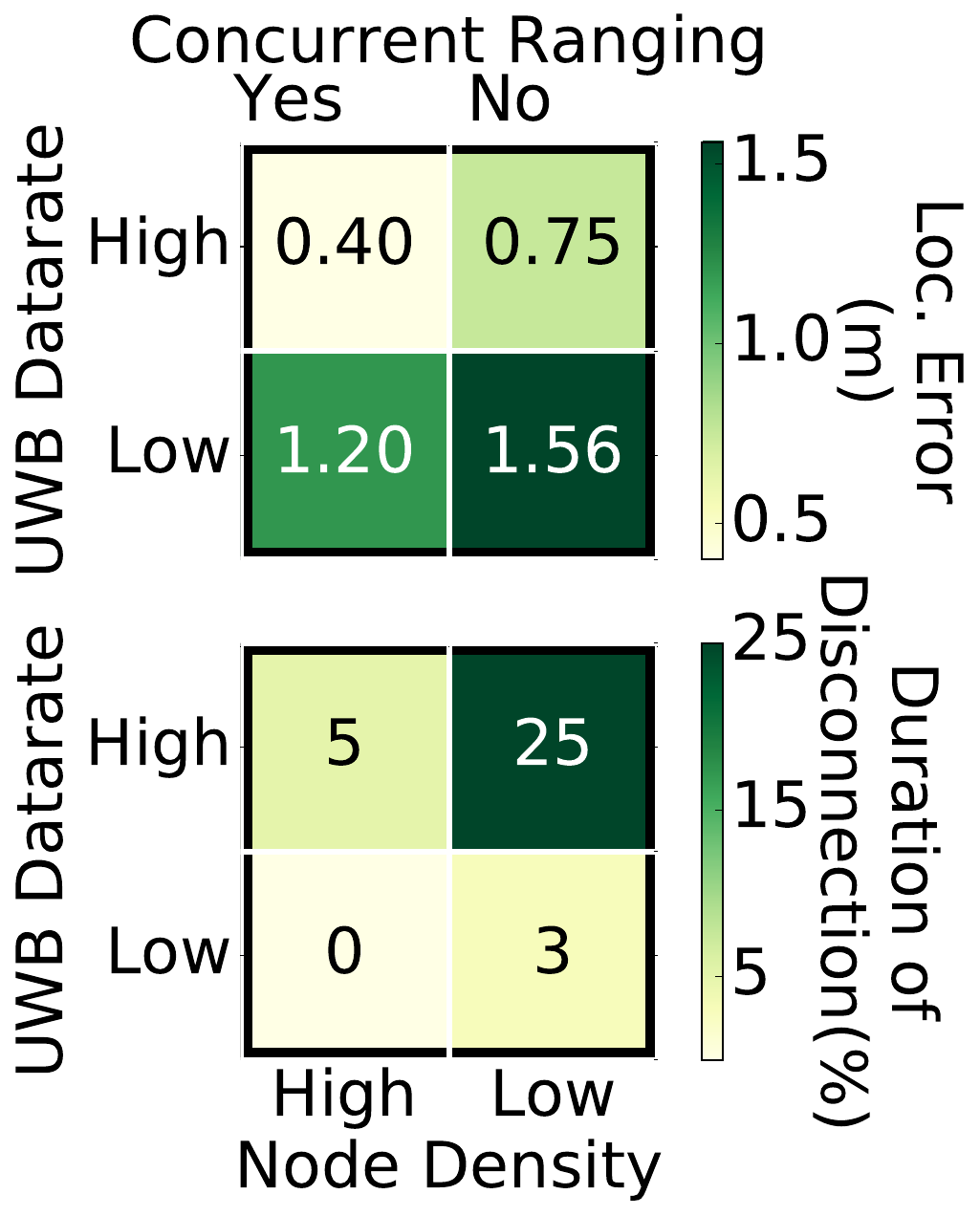}
	\label{fig:concurrent-ranging}
}
\caption{(a) Ranging error for different modes of UWB, (b) Loc. accuracy \& node dis-connectivity  for concurrent ranging, UWB datarates \& node densities}
\label{fig:concurrent}
\end{figure}

\noindent\underline{\textit{Concurrent Ranging:}} As seen in Fig.~\ref{fig:concurrent}(b), \system's aggregated and concurrent ranging directly contributes to a larger ranging budget and hence localization accuracy. \system's tag supports two data rates (low, 100\,Kbps and high, 6.8\,Mbps, Fig.~\ref{fig:concurrent}(a)). The high data rate results in a lower latency per ranging (2\,ms vs 8\,ms for low data rate case). While it restricts the maximum communication range (about 10 m) and allows for more concurrent ranging, it also requires a higher node density to ensure a reasonably connected topology. This is reflected in the results presented in Fig.~\ref{fig:concurrent}.  

\begin{figure}[!htb]
\centering
\subfloat[]
{
	\includegraphics[width=0.4\linewidth, height =90pt]{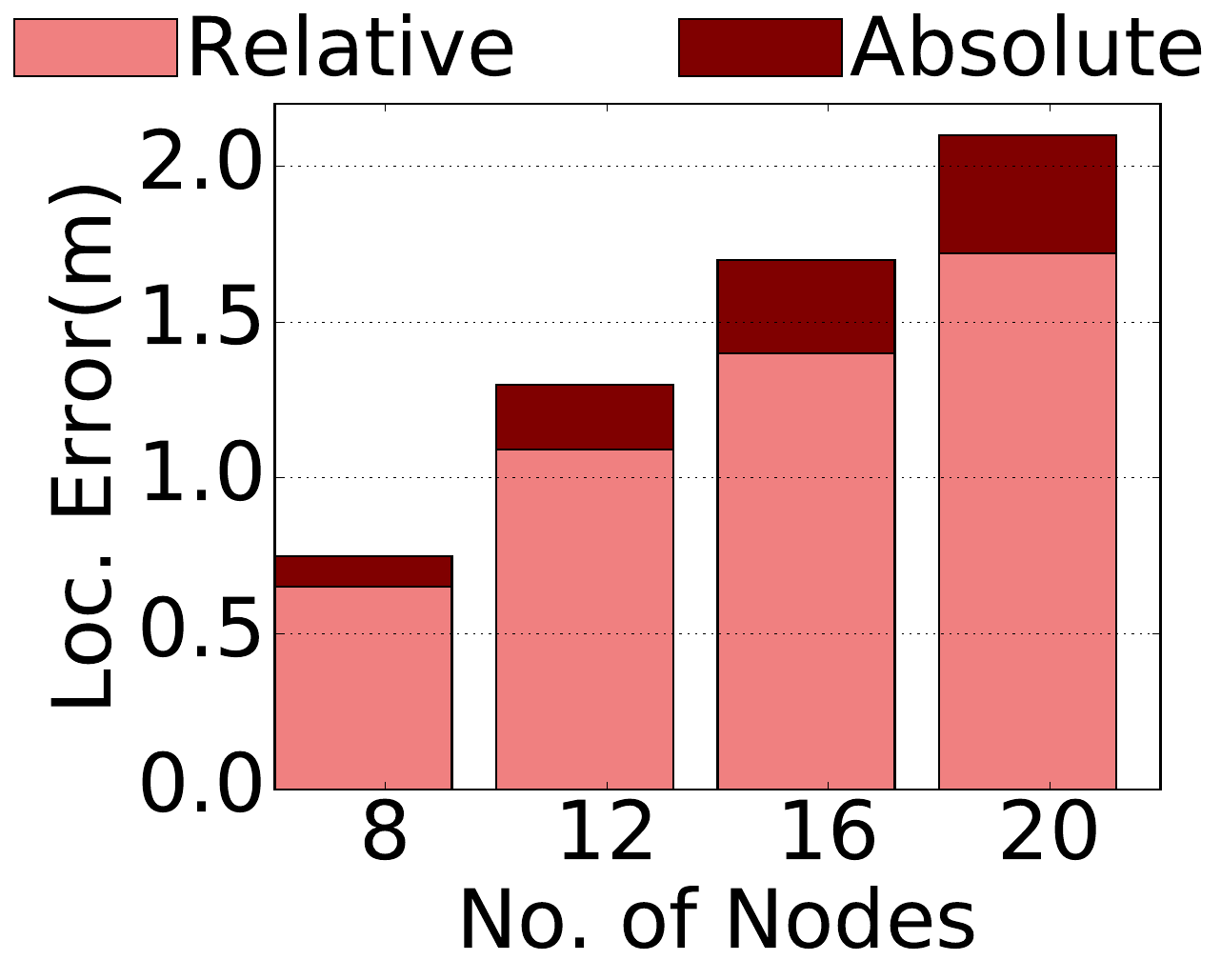}
	\label{fig:abs-vs-rel}
}
\subfloat[]
{
	\includegraphics[width=0.45\linewidth, height=90pt]{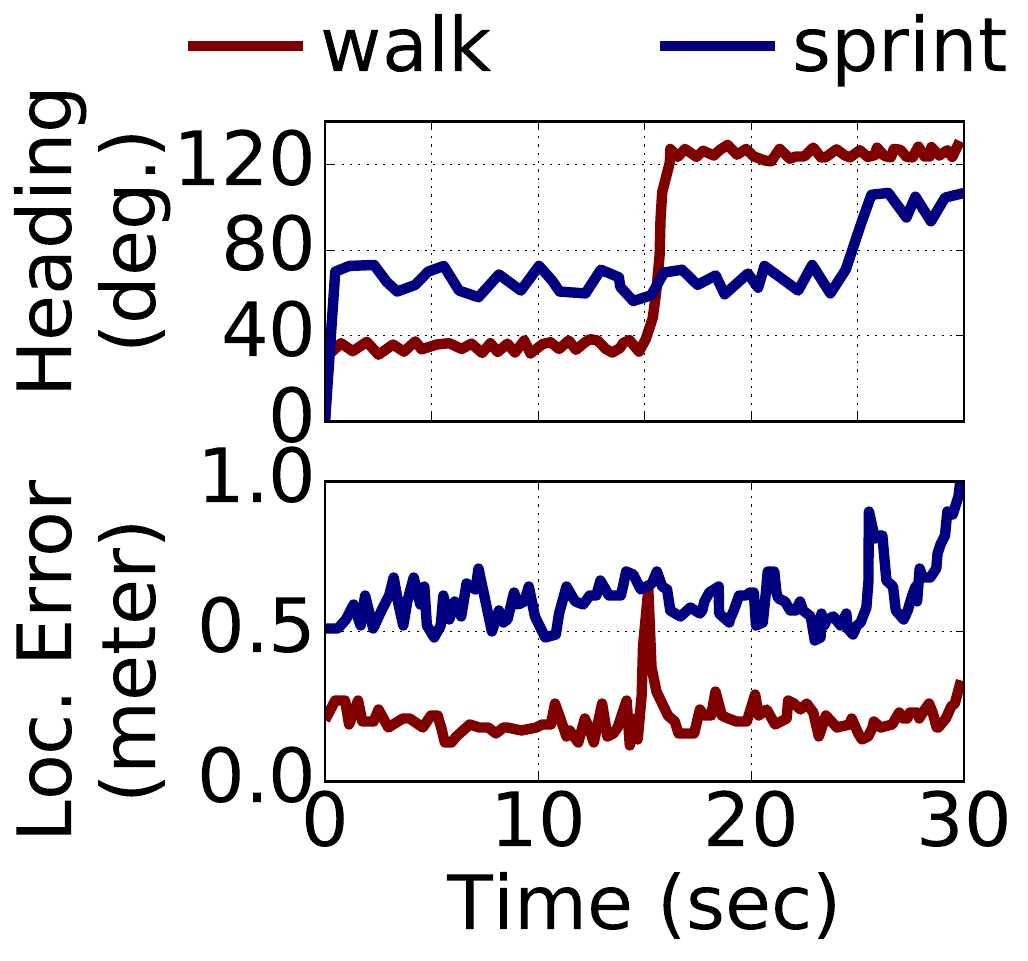}
	\label{fig:heading-accuracy}
}
\caption{(a) Accuracy of absolute Localization, (b) Impact of heading error}
\label{fig:abs}
\end{figure}

\begin{figure}[!htb]
\centering
\subfloat[]
{
	\includegraphics[width=0.5\linewidth]{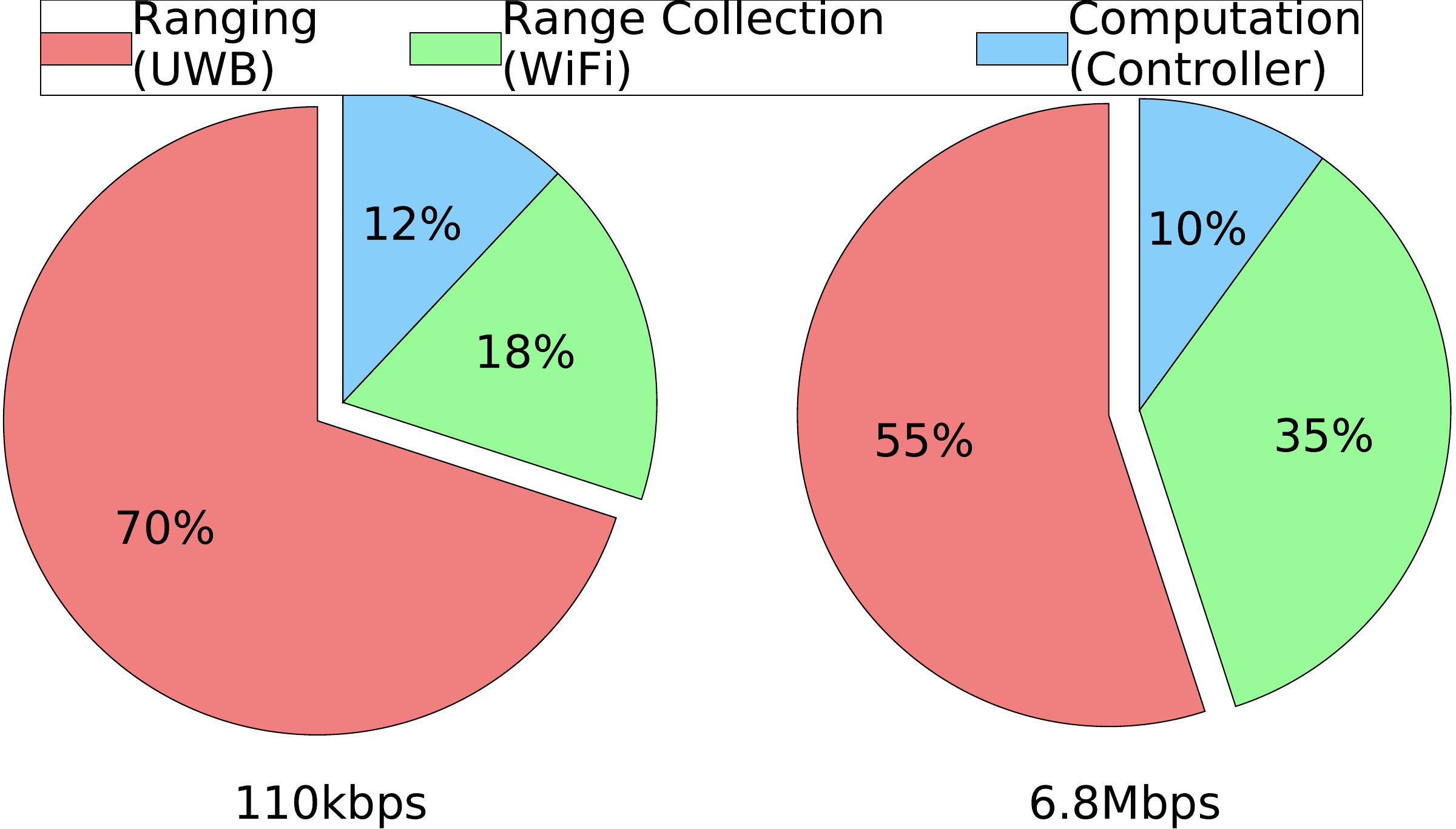}
	\label{fig:latency-effect-a}
}
\subfloat[]
{
	\includegraphics[width=0.45\linewidth]{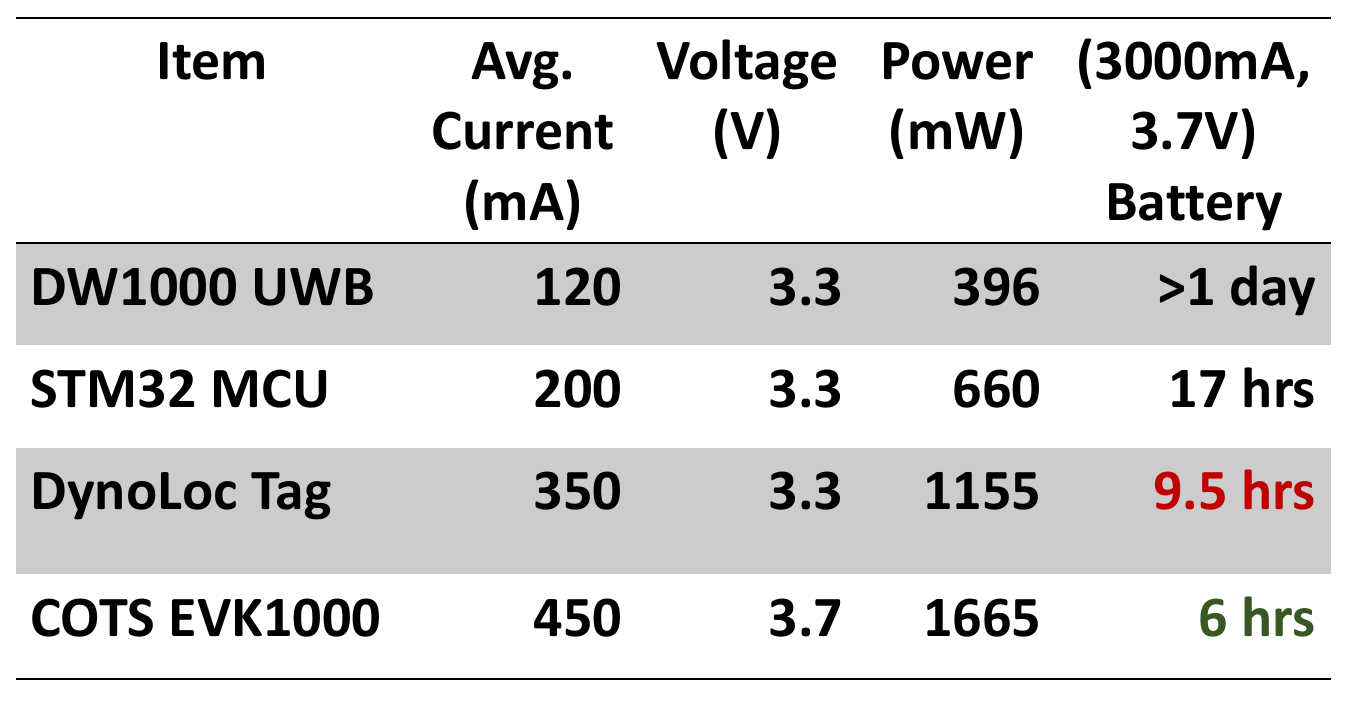}
	\label{fig:latency-effect-b}
}
\caption{(a) Breakdown of latency components (b) Power benchmarks for various system components}
\label{fig:latency-effect}
\end{figure}

\noindent\underline{\textit{Absolute Localization:}} Conversion of relative to absolute localization incurs an additional error of 10--20\%, as shown in Fig.~\ref{fig:abs-vs-rel}. As expected, it increases as the network size grows larger. Fig.~\ref{fig:heading-accuracy} shows how the value of heading is tracked over time as the user moves (walk and sprint). While noisy heading values lead to higher additional errors, this is still restricted to just 40-60\,cm even for a user sprinting at 2\,m/s. 
\\\noindent \underline{\textit{End-to-End System Latency:}} Fig.~\ref{fig:latency-effect-a} breaks down \system's overall latency broadly into three categories. In table~\ref{fig:latency-effect-b}, we present some battery life benchmarks for different components of the system. 
%

%

%
%
\section{Applications}
%
\label{sec:applications}
\system enables a host of location-based applications that benefit from minimal setup time, preferably without any infrastructure deployment, realtime support, and adaptation to an unknown environment. We demonstrate two such use cases that would benefit from \system. 

\subsection{Multiplayer AR/VR Gaming}
Today's multiplayer AR/VR gaming systems, do not support features that have bearing to players' location in the physical coordinate space. Given 
the growing demand for `location-based entertainment'~\cite{locentertainment}, 
recent solutions make use of visual SLAM (using VIO) to localize players with respect  to their individual reference frames~\cite{game2}. However, for a collaborative multiplayer setting, a global coordinate system is essential. 
This is accomplished today using either visual markers or anchors that do not offer a smooth multiplayer experience, or expensive/extensive installation of IR cameras and laser tags that do not offer a cost-effective, on-demand deployment (e.g. consumer homes).

\begin{figure}[!htb]
	\centering
	\includegraphics[width=0.9\linewidth]{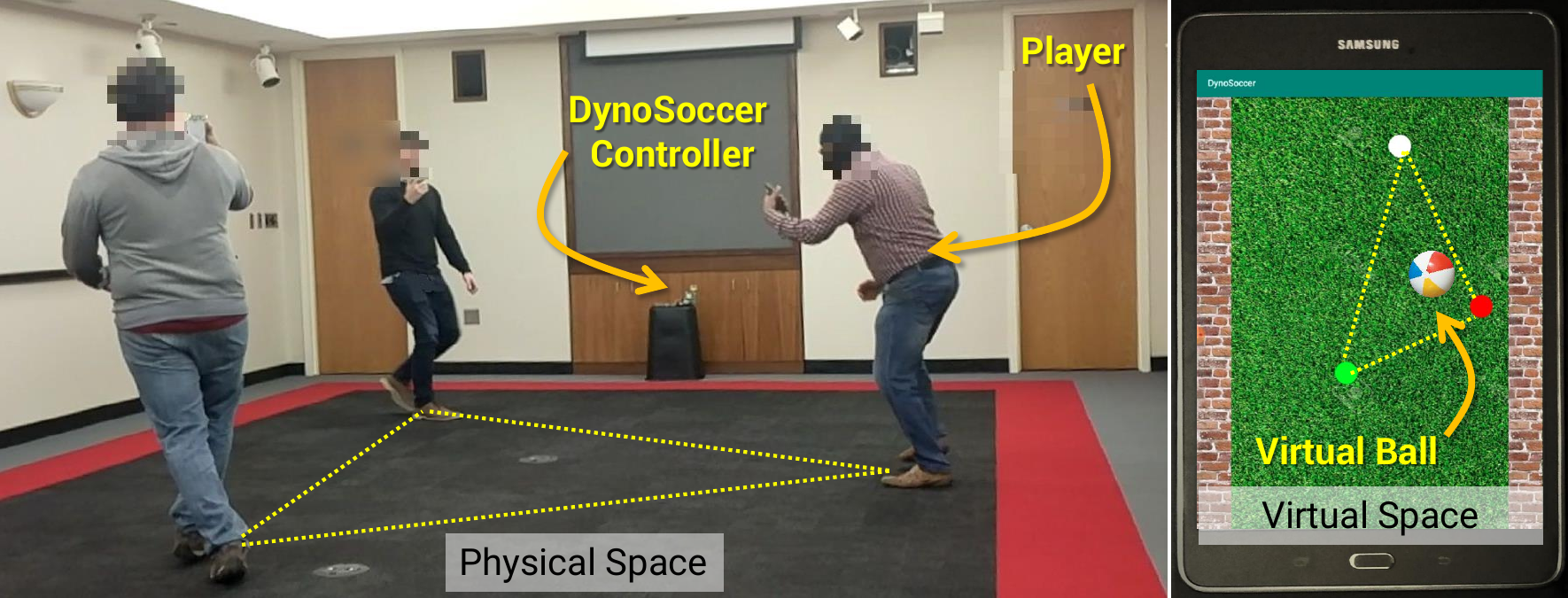}
	\includegraphics[width=0.55\linewidth, height=75pt]{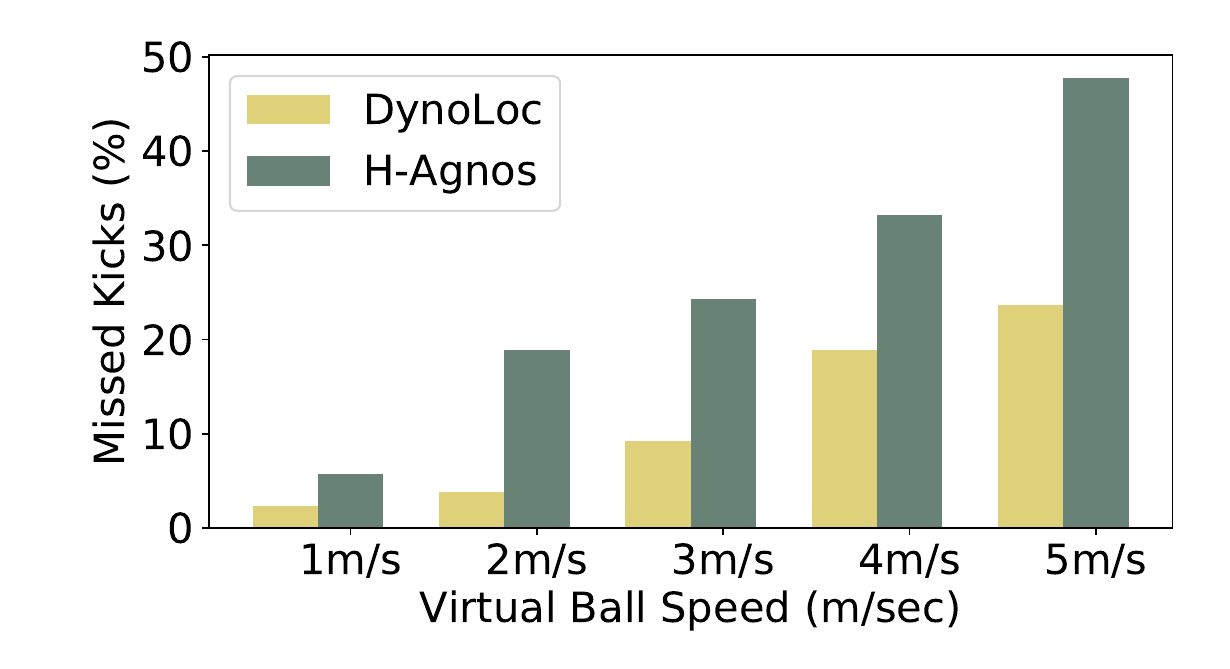}
	\includegraphics[width=0.43\linewidth, height=73pt]{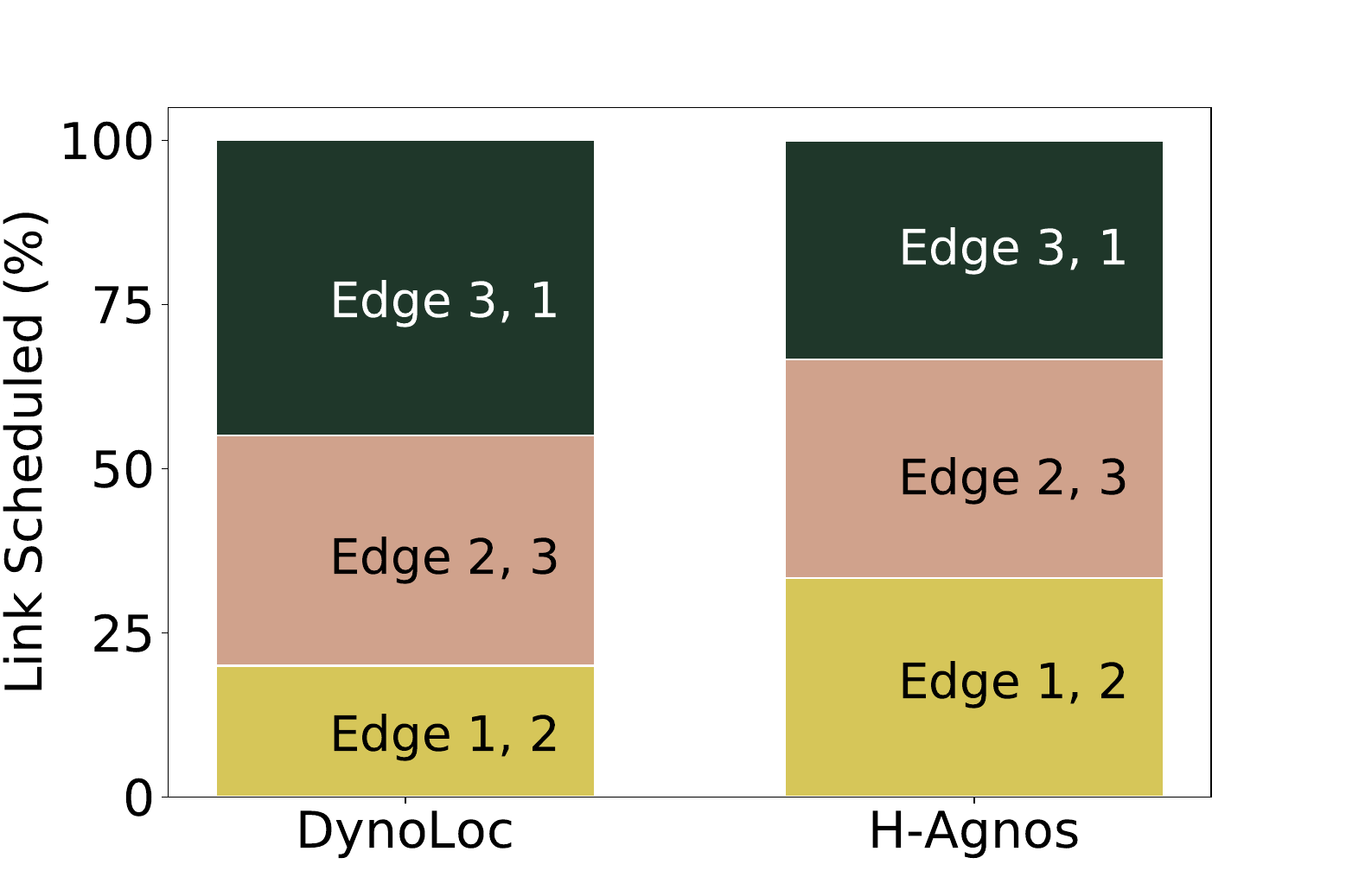}
	\caption{ {\em Top:} \system enables location-based multiplayer gaming. {\em Bottom:}  Mobility-aware range scheduling distributes time slots heterogeneously across nodes, thus improves overall interactivity of the game.}
	\vspace{-0.2 cm}
	\label{fig:app-1}
\end{figure}

We create a simple Android based multiplayer VR game called \texttt{DynoSoccer} to demonstrate the value that \system can bring to such gaming systems. \texttt{DynoSoccer} transforms ordinary physical spaces, like a living room 
that is not particularly well-lit or textured
, into a gaming arena, where players can interact with each other based on their {\em real} physical locations. The game consists of a virtual ball that is bounced around by the players. Each player needs to adjust their position to be in the proximity of the ball in order to `kick' it. However, this requires the system to be responsive to the players' and the ball's movement, otherwise resulting in a `missed kick'. We show (Fig.~\ref{fig:app-1}) that even for high ball speeds, \system results in 50\% to 90\% less `missed kicks' compared to the \texttt{H-Agnos} baseline, increasing the VR experience significantly. \system adaptively schedules the relevant ranges (compared to round-robin in \texttt{H-Agnos}) in the topology  based on the mobility of individual players.
We plan to fuse VIO with \system to overcome VIO's challenges in less-favorable visual environments (Fig.~\ref{fig:vio-indoor-exp}), especially in the multi-player context.
 %
%
\subsection{Active Shooter Scenario}

\begin{figure}[!htb]
	\centering
	\includegraphics[width=0.96\linewidth]{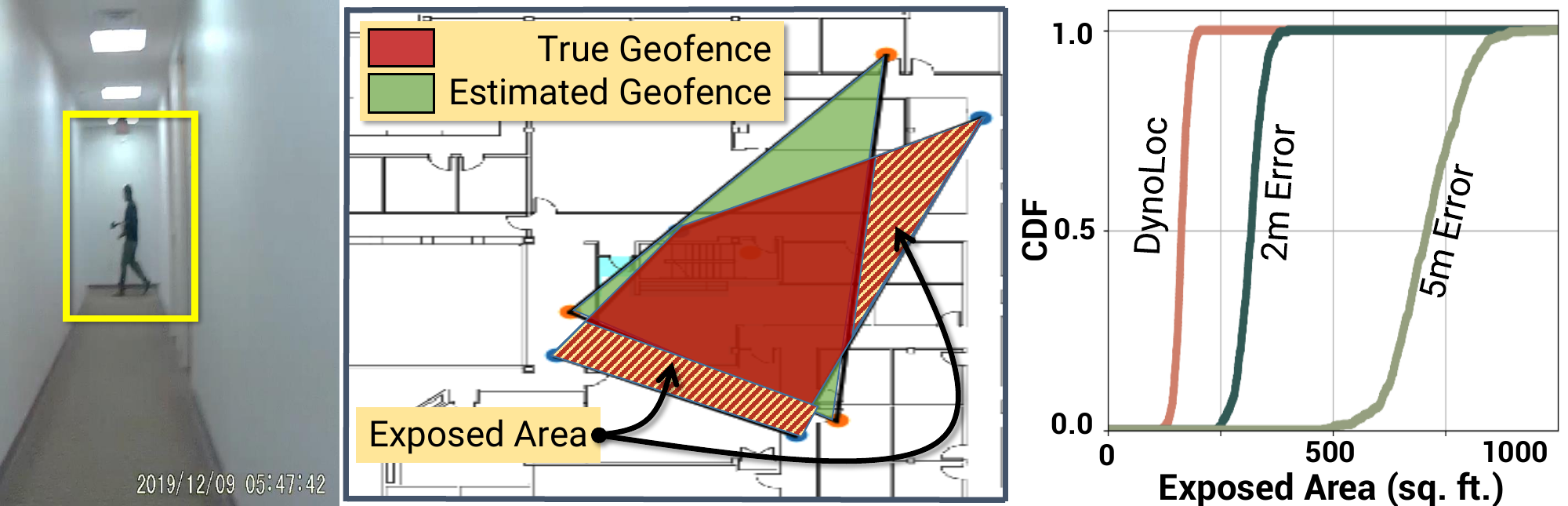}
	\caption{\system enables real time geofencing in an active shooter scenario helping in safe evacuation.}
	\label{fig:app-2}
\end{figure}

We demonstrate in Fig.~\ref{fig:app-2} how \system can create a realtime geofence for safe evacuation of trapped victims in a chaotic situation like spotting an active shooter. 
A person mimicking an active shooter runs in a specified path. Four volunteers (unaware of the shooter's path) equipped with \system tags and body camera scout the general area (corridors, hallways etc.). If the shooter is detected in the video frame, we mark the respective location as unsafe and update the geofence (a polygon connecting the unsafe points). We show in Fig.~\ref{fig:app-2} ({\em right}), how localization error can lead to inaccurate geofencing resulting in `exposed areas' or zones that are potentially dangerous but marked safe. Even a 2\,m median localization error can lead to a few hundred sq. ft. of exposed area, compared to the \system's limited exposure.
\section{Conclusion}
%
We introduced the problem of latency-bounded infrastructure free localization that is central to several dynamic indoor applications. Towards addressing the fundamental tradeoff between latency and localization accuracy that arises in these problems, we presented the design and practical  realization of our \system system. 
Through various design innovations, \system has demonstrated its ability to accurately track a large network of highly mobile entities, without any infrastructure support, in real-world firefighters' drill, as well as applications of multi-player AR gaming, and active shooter tracking.
%
%
\bibliographystyle{IEEEtran}
\bibliography{newref,ref,mobicom,postmobicom}

\clearpage
 \appendices
\section{Aggregate Ranging}
\begin{figure}[!htb]
  \begin{center}
    \includegraphics[width=0.5\textwidth]{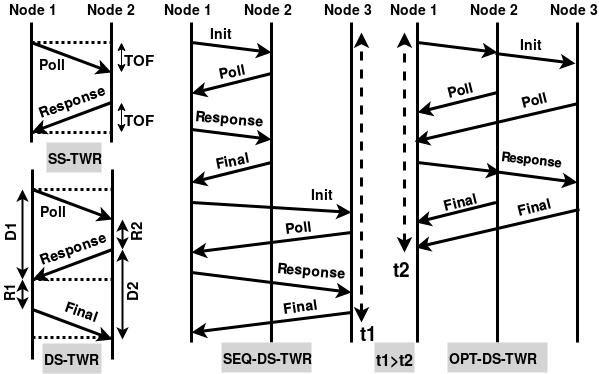}
  \end{center}
  \caption{Two-way ranging (a): Single-sided (SS) (b) Double-sided (DS) (c) Sequential Double-sided (SEQ-DS) and (d) Optimized Double-sided (OPT-DS) Two Way Ranging (TWR). }
  \label{fig:twr}
\end{figure}
UWB employs a two-way ranging (TWR) mechanism \cite{sahinoglu2006ranging} , standardized in IEEE 802.15.4, to estimate the distance between a node pair based on Time of Flight (TOF). Further, each node pair needs to be separately ranged following a strict TDMA schedule to avoid interference.
This leads to a sequential ranging for the desired node pairs, consuming significant latency and hindering scalability.
TWR involves exchanging 4 messages for a single range estimation. See Figure \ref{fig:twr}. \system instruments an aggregated version of TWR where an {\em initiator node} sends a broadcast \texttt{INIT} message that also contains \texttt{ID}s of other nodes that it wants to range with in tandem. The receiving nodes take turns individually to send a \texttt{POLL} message, which is followed by a broadcast \texttt{RESPONSE} message from the initiator node. On receiving the \texttt{RESPONSE} message, the nodes take turn once again to send a \texttt{FINAL} message. Based on the timestamps in the above messages the initiator node calculates the distance estimates to individual nodes. Thus, our aggregated optimized TWR (OPT-DS-TWR in Fig. \ref{fig:twr}) can range N nodes in (2N+2) time slots compared to the 4N slots for its sequential counterpart. Given that each transmission slot time can be 10\,ms, this substantial saving in turn allows for {\em twice} as many links to be ranged with the same latency budget. 

\section{Concurrent Ranging}
	We use concurrent ranging to increase ranging budget and hence support more nodes for a given refresh rate. 
While, the UWB protocol employs TDMA to avoid wireless contention, it does not account for spatial reuse, wherein node pairs outside each others interference domain can be operated simultaneously. \system can easily compute such link concurrency information based on the neighborhood information for each node as discussed earlier. Hence, a set of initiator nodes and their corresponding ranging nodes are logically partitioned into non-interfering groups, with every group containing an initiator node. 
Each of the initiator nodes executes the aggregated TWR process concurrently to range relevant nodes in its group. Such concurrent ranging, that makes use of spatial reuse, provides a significant increase in the ranging budget that is beneficial for large node topologies.

\system employs the above ranging protocol to collect ranges for links in its ranging topology. 
Such optimizations also result in a significantly larger ranging budget (within the application's refresh rate)  for use by its edge selection component that in turn contributes to a larger localization accuracy.

\section{EDM Completion Algorithm}
	Following the discussion in Section 3.4(A), Algorithm \ref{algo:edm-completion} describes the steps of EDM completion.
\begin{algorithm}[htb!]
	\caption{EDM Completion Algorithm}
	\label{algo:edm-completion}
	\begin{algorithmic}[1]
		\scriptsize
		%
		%
		\STATE Initialize EDM $D$ with measured ranges
		\STATE Compute node locations using multilateration on the rigid graph
		\STATE Fill in the missing values in $D$ using node locations
		\STATE Initialize EDM $D_{cur}$ using node locations
		\WHILE{$\Delta D = |D - D_{cur}|$ is significant}
			\STATE Update location of node i, j for each edge (i, j) in proportion to the $\Delta D(i, j)$
			\STATE Recompute $D_{cur}$ from node locations
			\STATE Fill in missing values in $D$ from node locations
		\ENDWHILE
		\STATE Output Complete EDM $D$
	\end{algorithmic}
\end{algorithm}

\section{Absolute Localization}
\system needs to transform the localization solution from a  relative coordinate system to a target absolute coordinate system. 
By translating, rotating and flipping (if needed) the relative coordinate axis, the absolute localization solution can be determined in one of two 
ways.
%
%
\begin{figure}[!htb]
  \begin{center}
    \includegraphics[width=0.18\textwidth]{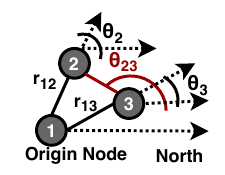}
  \end{center}
  \caption{Finding the slope of an edge}
  \label{fig:abs-loc-rot}
\end{figure}

\mypara{Fixed reference node.} One of the static nodes in the topology is made to serve as the reference node. This could be a node that is placed on an entrance door (e.g. when responders enter a building). 

\noindent \textit{\underline{Translation}}: The reference node becomes the origin of the absolute coordinate axis by translating the relative coordinate axis accordingly.
%

\noindent \textit{\underline{Rotation and Flip}}: \system rotates the translated coordinates by  an angle $\theta$, determined as follows. Assume the positive x-axis of the absolute coordinate system to be aligned with earth's magnetic North direction. The heading from the IMU gives the angle its axis makes with this direction ($\theta_2$ \& $\theta_3$ in Figure \ref{fig:abs-loc-rot}). When the nodes are in motion, the heading and the direction of motion can be assumed to be the same (contrast this with sideways motion, which is however, atypical). Using inter-node distances ($r_{12}$ \& $r_{13}$ in Figure \ref{fig:abs-loc-rot}) and the headings, the slope of an edge (e.g. $\theta_{23}$ for edge (2,3)) with the north direction and its slope (say $\theta^{'}_{23}$) in the relative coordinate are determined. The required rotation angle is then the difference between these two slopes ( i.e $\theta = \theta_{23} - \theta^{'}_{23}$). \system  also corrects the flip by considering the vertex-ordering of the nodes with respect to the centroid  of the coordinates in a clockwise direction. If the two orderings (corresponding to the relative and absolute coordinates) do not match, the relative coordinates are flipped across the x-axis to derive the final absolute coordinates of the nodes.

\begin{figure}[!htb]
  \begin{center}
    \includegraphics[width=0.8\linewidth]{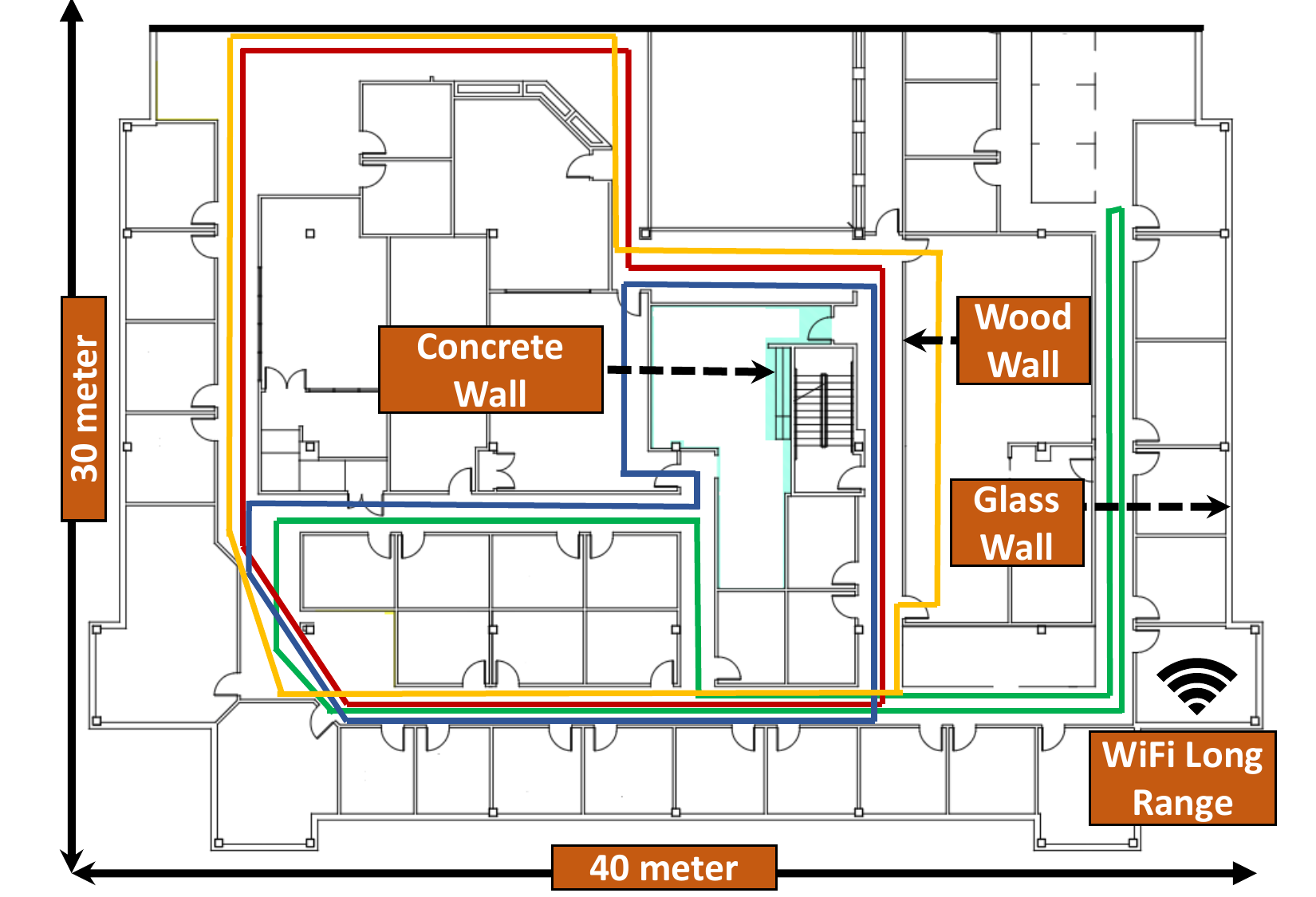}
  \end{center}
  \caption{\sfrtxt{Floorplan with the a subset of the loops traversed by the mobile nodes in the controlled experiments}}
  \label{fig:exp-paths}
\end{figure}

\mypara{Floormap.} If the floormap is available, \system first applies the rotation and flip correction. Additionally, the trajectory traversed by each node is saved. To fix the translation, \system tries to place the joint set of trajectories within the bounding box of the floormap by finding a location such that none of the trajectories  cross any walls and the distances among various points across the trajectories  are satisfied according to the measured ranges~\cite{xie2016xd}. Note that this approach does not require any fixed static node as a reference/origin. \sfrtxt{The floormap and the paths used in the controlled experiment is illustrated in Fig.\ref{fig:exp-paths}.}
\section{In The Wild Experiment}
\begin{figure}[!htb]
	\centering
	\includegraphics[width=0.71\linewidth]{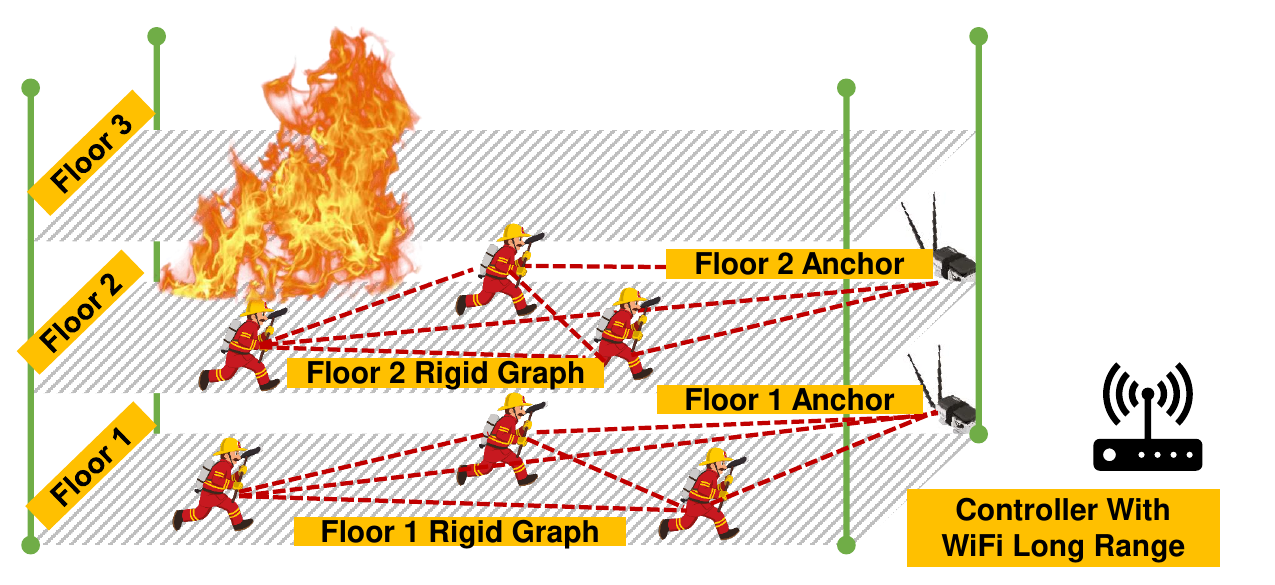}
	\includegraphics[width=0.28\linewidth]{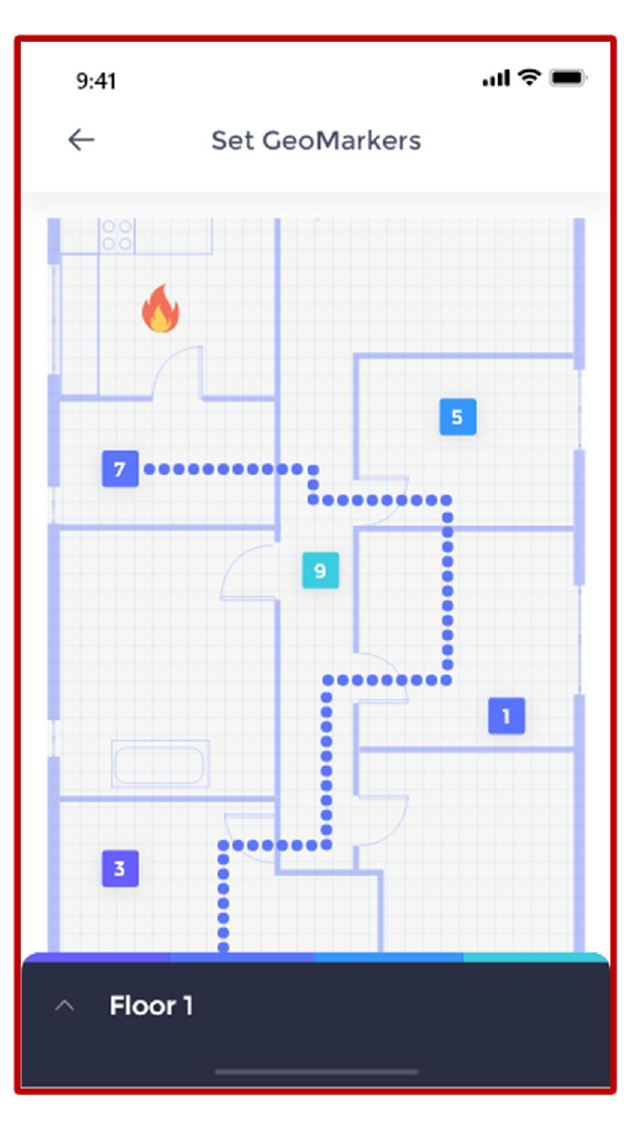}
	\caption{
		\sfrtxt{(a) \system{} operation across multiple floors, (b) GUI of \system Android app showing node locations on input floormap with breadcrumb and geo-tagging features.}
	}
	\label{fig:vertical-detection}
\end{figure}

	\sfrtxt{
As shown in Fig.\ref{fig:vertical-detection}(a), absolute localization is run by \system in each floor independently. To determine the the current floor, the reading from the pressure sensors (also available in the smart-phones) in conjunction with the available UWB connectivity information are used by \system{}. Ideally, nodes from each floor form a disjoint connected graph. However, due to shared vertical space among floors i.e. staircase, a node may have links with nodes on more than one floor. In those cases, floor of the node in question is determined by pressure sensors. To facilitate uncalibrated operation, during boot-up of \system{}, each node's initial pressure at the ground level is recorded and then continuously tracked for changes in shorter period  (actual floor change) and longer period (typically $\geq$20 minutes, which signifies atmospheric pressure change). This pressure sensor data is piggy-backed with the range data just like link quality and mobility metric information.
}

%
%
\end{document}